\documentclass[12pt,onecolumn]{IEEEtran}
\usepackage{cite}
\usepackage[cmex10]{amsmath}
\usepackage{array}
\usepackage{amssymb}
\usepackage{amsfonts}
\usepackage{subeqnarray}
\usepackage{cases}
\usepackage{mathrsfs}
\usepackage{graphicx}
\usepackage{threeparttable}
\usepackage{acronym}
\usepackage{caption}
\usepackage{subcaption}
\usepackage{dsfont}
\usepackage{lineno}
\begin{document}
\title{CFD results calibration from sparse sensor observations with a case study for indoor thermal map}

\author{\IEEEauthorblockN{Chaoyang Jiang, Yeng Chai Soh$^*$,  Hua Li, Mustafa K. Masood, Zhe Wei, Xiaoli Zhou, and Deqing Zhai}

\thanks{C. Jiang, Y. C. Soh, M. K. Masood, Z. Wei, and D. Zhai are with the School of Electrical and Electronic Engineering, Nanyang Technological University, 639798 Singapore (e-mail: chaoyangjiang@hotmail.com; eycsoh@ntu.edu.sg).

H. Li and X. Zhou are with the school of Mechanical and Aerospace Engineering, Nanyang Technological University, 639798 Singapore (e-mail: lihua@ntu.edu.sg.).}
}

\maketitle

\begin{abstract}
Current CFD calibration work has mainly focused on the CFD model calibration. However no known work has considered the calibration of the CFD results. In this paper, we take inspiration from the image editing problem to develop a methodology to calibrate CFD simulation results based on sparse sensor observations. We formulate the calibration of CFD results as an optimization problem. The cost function consists of two terms. One term guarantees a good local adjustment of the simulation results based on the sparse sensor observations. The other term transmits the adjustment from local regions around sensing locations to the global domain. The proposed method can enhance the CFD simulation results while preserving the overall original profile. An experiment in an air-conditioned room was implemented to verify the effectiveness of the proposed method. In the experiment,  four sensor observations were used to calibrate a simulated thermal map with $167\times 365$ data points. The experimental results show that the proposed method is effective and practical.
\end{abstract}

\IEEEpeerreviewmaketitle

\section{Introduction}

The estimation of the indoor physical field (e.g., thermal map, airflow pattern, and pressure field) is an important problem in air-conditioning system design and indoor architectural-design. In recent years, the computational fluid dynamics (CFD) simulation has been widely used in building design to predict the thermal map, pollution dispersion, and ventilation performance \cite{zhai2006application, chen2009ventilation}. The accuracy of the CFD simulation strongly depends on the setting of boundary conditions and CFD models \cite{srebric2008cfd}.
In practice, the real boundary conditions of the rooms or buildings are unavailable and the idealized version is used instead. Even though CFD modelling has been extensively researched, the reliability of the results remains a key concern when performing CFD simulations.

Currently, in order to produce reliable simulation results, we need to properly set the boundary conditions and adjust the computational model (numerical and/or physical models) input parameters to amend the agreement between the simulated results and the corresponding experimental data \cite{hajdukiewicz2013formal, guillas2014bayesian, kajero2016kriging}. The current CFD model calibration work has mainly focused on minimizing the mismatch between simulated and experimental results.  Tens to hundreds of simulation runs are required for the calibration process of finding acceptable model parameters. However, even a well calibrated CFD model cannot remove the gap between simulated and experimental data, which implies that the results of calibrated CFD models still have room for further improvement.

CFD simulations can map the sampled spatial domain to the airflow-field data. It can provide the global information of airflow fields of interest, which is rather expensive or even impossible to obtain from real experimental observations. On the other hand, sensors (e.g., thermocouples) can only practically provide sparse and isolated observations, but the observations are usually much more accurate than the simulation results.  Can we calibrate/correct the simulated physical fields with sparse sensor observations? In this paper, we develop a methodology to solve this problem. To the best of our knowledge, no reported publication has focused on this problem.

The recent work \cite{hajdukiewicz2013formal, guillas2014bayesian, kajero2016kriging} and the references therein mainly focused on the calibration of a CFD model. In this paper, we directly calibrate the results of CFD simulations.
The proposed methodology can be applied for general CFD simulations and be a valuable additional step of the standard CFD simulation procedure.

\begin{figure}[htb]
    \centering
    \begin{subfigure}[b]{0.25\textwidth}
        \centering
        \includegraphics[width=\textwidth]{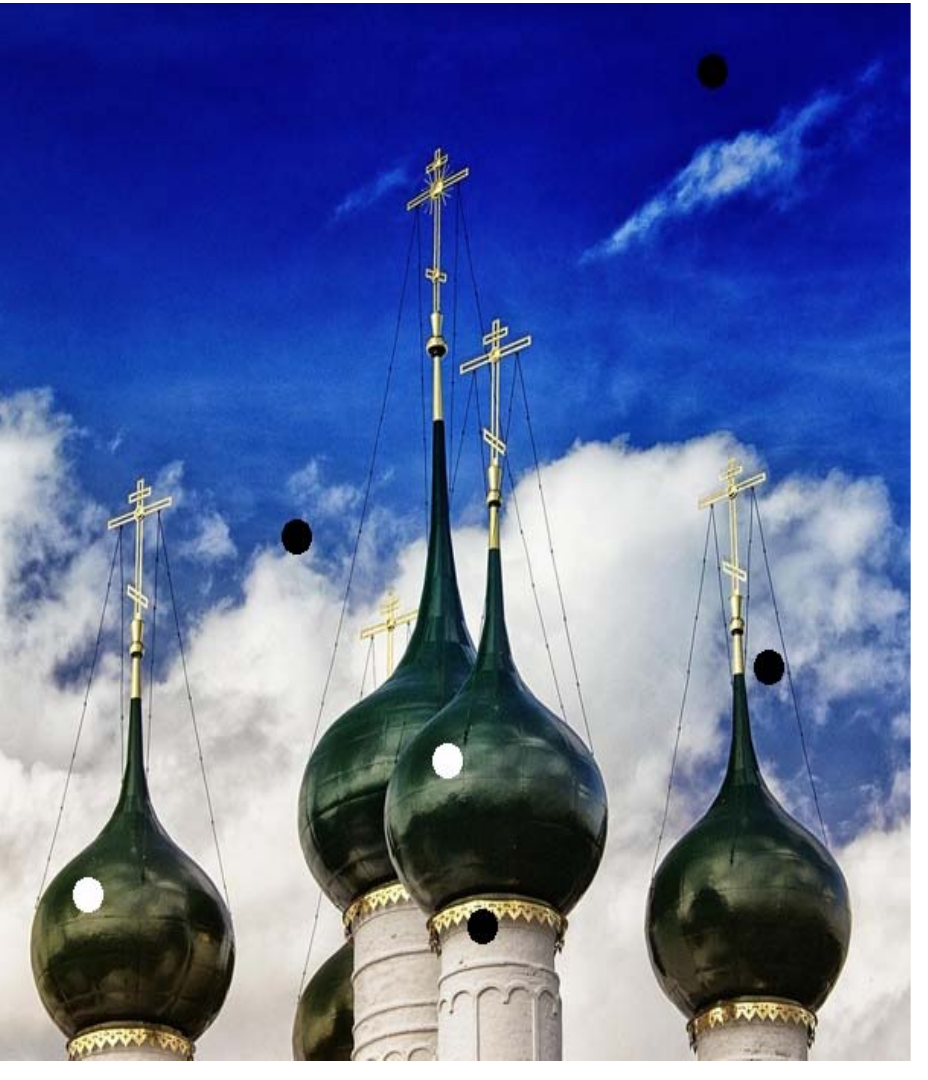}
    \end{subfigure}%
    \begin{subfigure}[b]{0.25\textwidth}
        \centering
        \includegraphics[width=\textwidth]{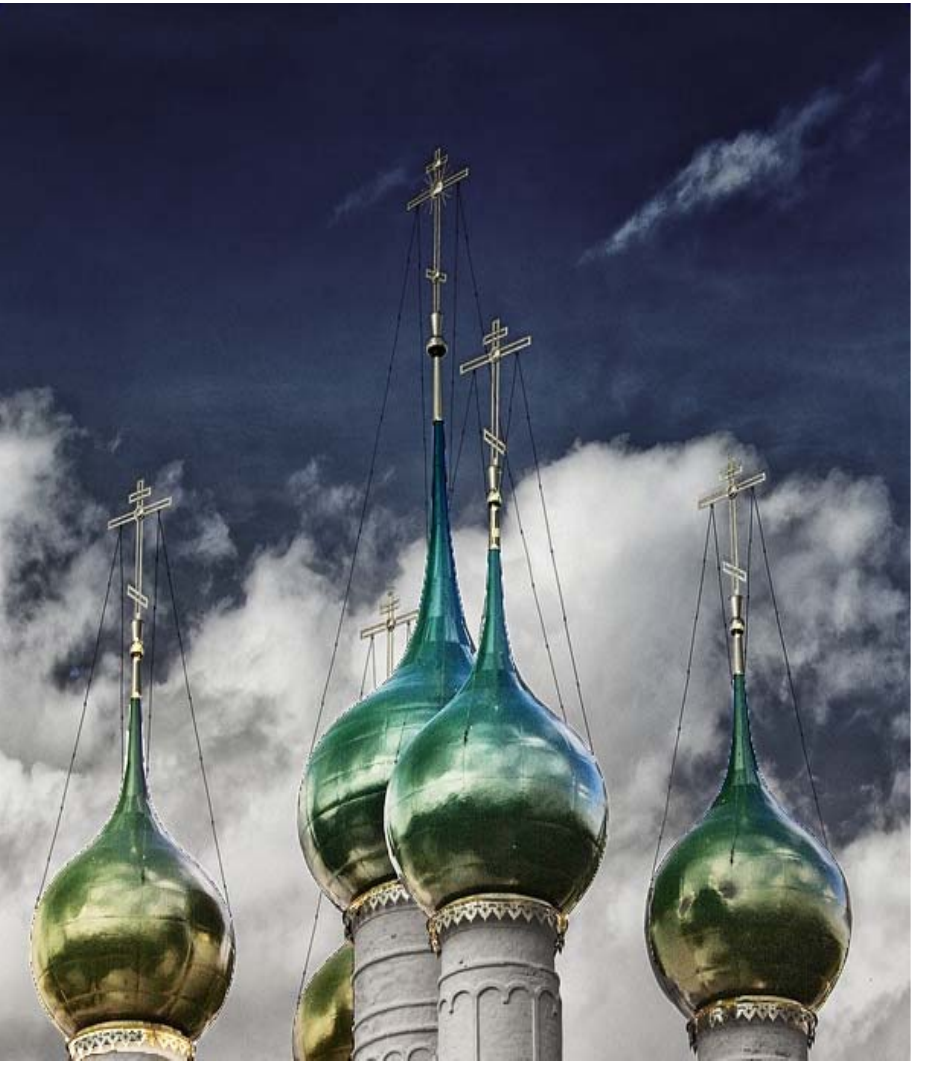}
    \end{subfigure}
 \caption{An simple example of image editing \cite{xu2013sparse}. The dots in the left image are used as control samples to edit the input image (i.e., left image). The right image is the results of the image editing.}
    \label{Fig:ImageEditing}
\end{figure}

In this work, we propose a methodology to calibrate the physical fields obtained from CFD simulations using sparse sensor observations. Since sensor observations are much more accurate than simulation results, we ignore the difference between sensor observations and the real values and use them as the ground truth. In addition, we assume that the simulation results can provide the rough global profile of the real physical fields, i.e., the profile of the simulated physical field is similar to that of the real physical field. The authors have taken inspiration from the image editing technique \cite{an2008appprop, xu2013sparse}.
As shown in Fig. \ref{Fig:ImageEditing}, an image can be globally edited with sparse control samples. The right image is fused by the original image and the sparse control samples. The result of image editing reserves the content of the original image but with the style of the sparse control samples. For the calibration problem of simulated physical field, we can view the sparse sensor observations as the control samples. The calibration problem can be viewed as a simulated physical field editing problem.

We formulated the simulated physical field calibration as an optimization problem. The objective function is a sum of two parts. One part is responsible to narrow the difference between the sensor observations and the simulation results, while the other part guarantees the similarity between the simulated results and the calibrated results. The optimization problem is solved by a numerical method. The proposed method was tested in an air-conditioned room. With the help of four sensor observations, indoor thermal map obtained from CFD simulation was well calibrated.

We summarize the contributions of this work as: 1) we proposed a formulation for the calibration of simulated physical fields from sparse sensor observations; 2) we formulated the proposed calibration problem as an optimization problem; 3) we provided a computationally efficient method to solve the optimization problem, and 4) we verified the effectiveness of the proposed methodology in an air-conditioned room.

The rest of this paper is organized as follows. In Section II, we formulate the CFD-result calibration problem. In Section III, we detailed the proposed methodology in which we first formulate the calibration problem as an optimization problem, then present a numerical method to solve it, and finally we provide a computationally efficient approximated solution. In Section IV, we calibrated a thermal map of an air-conditioned room to verify the effectiveness of the proposed method. The concluding remarks are given in Section VI.

\section{Problem statement}

We denote a real physical field (e.g., thermal map, airflow pattern, and pressure field) of interest by $f(\mathbf{x})$, where $\mathbf{x}=(x,y,x)\in\Omega$ with $\Omega$ being the domain of interest. The physical field obtained from CFD simulation is a numerical solution, which can be simply interpolated to be a continuous one. We denote a continuous physical field interpolated from a CFD solution by $f_c(\mathbf{x})$, which can be described as
\begin{equation} \label{eq:fcfd}
  f_c(\mathbf{x})=f(\mathbf{x})+v(\mathbf{x}), \quad \mathbf{x}\in \Omega
\end{equation}
where $v(\mathbf{x})$ represents the error of the simulated physical field. We assume that $\max_{\mathbf{x}\in \Omega}v(\mathbf{x})\leq v_{\mathrm{max}}$, and $v_{\mathrm{max}}$ is a small value compared with the magnitude of $f(\mathbf{x})$.

We denote one of the sensor observations of the physical field by
\begin{equation}\label{eq:sensorobservation}
  s_k = f(\mathbf{x}_k)+n_k, \quad \mathbf{x}_k = (x_k, y_k, z_k)\in\Omega_s\subset\Omega, k=1,2,\ldots m
\end{equation}
where $n_k$ is the measurement noise of the $k$-th observation, $\Omega_s$ is the set of all sensing locations, and $m$
is the number of available sensor observations. Without loss of generality, we assume that $|n_k|\ll |v(\mathbf{x}_k)|$ for all $1\leq k\leq m$.

In practice, $f_c(\mathbf{x}), \mathbf{x}\in \Omega$ and $s_k, 1\leq k\leq m$ are known. In what follows, we estimate $v(\mathbf{x}), \mathbf{x}\in \Omega$ based on $f_c(\mathbf{x}), \mathbf{x}\in\Omega$ and $s_k, 1\leq k\leq m$. With the estimated $v(\mathbf{x})$, we can easily correct the simulated physical field $f_c(\mathbf{x})$.

\section{The proposed methodology}

\subsection{Formulation of the optimization problem}

Since $|n_k|\ll |v(\mathbf{x}_k)|$ for all $1\leq k\leq m$, we ignore the measurement noise. We denote the error of the simulated physical field at the sensing locations by $e_k$, and from \eqref{eq:fcfd} and \eqref{eq:sensorobservation} we can obtain
\begin{equation*}
  e_k = f_c(\mathbf{x}_k)-s_k = v(\mathbf{x}_k)-n_k \approx v(\mathbf{x}_k), 1\leq k\leq m
\end{equation*}
In the remainder of this paper, we ignore the measurement noise and view $e_k$ as the real error of $f_c(\mathbf{x}_k)$.

To estimate the error of the simulated physical field $f_c(\mathbf{x})$, i.e., $v(\mathbf{x})$, we minimize the following cost function
\begin{equation}\label{eq:EnergyFun}
J_0=\int_{\Omega}\sum_{k=1}^{m}w(\mathbf{x},\mathbf{x}_k)[v(\mathbf{x})-e_k]^2\mathrm{d}\mathbf{x}  +\lambda\iint_{\Omega\times\Omega}w(\mathbf{x},\mathbf{y})[v(\mathbf{x})-v(\mathbf{y})]^2\mathrm{d}\mathbf{x}\mathrm{d}\mathbf{y}
\end{equation}
where
\begin{equation}\label{eq:wight}
w(\mathbf{x},\mathbf{y})=\exp(-\|f_c(\mathbf{x})-f_c(\mathbf{y})\|^2/\sigma_m)\exp(-\|\mathbf{x}-\mathbf{y}\|^2/\sigma_d)
\end{equation}
is a priori known weight, which can be directly calculated from the simulated physical field. Note that $w(\mathbf{x},\mathbf{y})$ represents the affinity between $v(\mathbf{x})$ and $v(\mathbf{y})$. Both $\sigma_m$ and $\sigma_d$ are positive variances. It is obvious that $w(\mathbf{x},\mathbf{y})\in[0,1]$ and that $w(\mathbf{x},\mathbf{y})$ is very small unless the locations $\mathbf{x}$ and $\mathbf{y}$ are near and their corresponding physical value $f_c(\mathbf{x})$ and $f_c(\mathbf{y})$ are similar.

Solving the minimization problem \eqref{eq:EnergyFun}, we can find an estimation $v(\mathbf{x})$, which we denote by $\hat{v}(\mathbf{x})$. With $\hat{v}(\mathbf{x})$, we can calibrate the simulated physical field $f_c(\mathbf{x})$, and the calibration result is
\begin{equation}\label{eq:cali}
  \hat{f}(\mathbf{x})=f_c(\mathbf{x})-\hat{v}(\mathbf{x})
\end{equation}

The cost function \eqref{eq:EnergyFun} is the summation of two parts. The first part is responsible for  the regions around sensing locations. In the region around $\mathbf{x}_k$, the value of $v(\mathbf{x})$ is forced to be similar to $e_k$, especially when $f_c(\mathbf{x})$ is similar to $f_c(\mathbf{x}_k)$. Consequently, in the region around $\mathbf{x}_k$,
the physical value is calibrated to be similar to the sensor observation $s_k$, especially for the subregion in which $f_c(\mathbf{x})$ is similar to $f_c(\mathbf{x}_k)$. In this way, in the region around $\mathbf{x}_k$, $\hat{f}(\mathbf{x})$ preserves the profile of $f_c(\mathbf{x})$ but the exact values are adjusted to be similar to the sensor observation $s_k$. The first term in \eqref{eq:EnergyFun} can guarantee a good local calibration for the simulated physical field, but it has insignificant influence on the regions far away from $\mathbf{x}_k$ for all $1\leq k\leq m$, in which $w(\mathbf{x},\mathbf{x}_k)\approx 0$.

On the other hand, the second term has a global influence, which can be explained from the following two perspectives: 1) the second term can transmit the adjustment from the local regions around the sensing locations to the whole physical field; 2) the second term can guarantee that in any local region, the gradient of $\hat{v}(\mathbf{x})$ is small if in that region the simulated values $f_c(\mathbf{x})$ are similar, which guarantees the profile of $\hat{f}(\mathbf{x})$ is similar to that of the simulated physical field $f_c(\mathbf{x})$.

The optimization problem is controlled by three parameters: $\lambda$, $\sigma_m$, and $\sigma_d$. $\lambda$, which we call the \emph{balance factor}, can balance the contribution of the two terms in \eqref{eq:EnergyFun}. A small $\lambda$ leads to good calibration in the local regions around sensing locations, while a large $\lambda$ can globally reduce the gradient of the adjustment at the cost of the calibration accuracy in the local regions around the sensing locations.

$w(\mathbf{x},\mathbf{y})$ is the product of two Gaussian functions. $\sigma_m$ and $\sigma_d$ control the rate of decay of the two gaussian functions, respectively.  $\sigma_m$, which we call the \emph{magnitude variance}, mainly depends on the dynamic range of the simulated physical field. If the local gradient of the physical field is large, a large $\sigma_m$ is proper. Otherwise the first Gaussian function in \eqref{eq:wight} will be very small which may dilute the influence of the second Gaussian function. On the other hand, if the thermal map has a low dynamic range, a relatively small $\sigma_m$ is proper.

$\sigma_d$, which we call the \emph{distance variance}, controls the scope of the local region around one of the sensing locations in which the physical data can be calibrated  by minimizing the first term in \eqref{eq:EnergyFun}. In other words, in the particular local region, the sensor observation have a significant influence on the physical field calibration.

\subsection{Numerical method to solve the optimization problem}
It is difficult for us to directly minimize $J_0$ in \eqref{eq:EnergyFun} to find the estimation of the error, i.e., $v(\mathbf{x}), \mathbf{x}\in\Omega$. Numerical methods can make the optimization problem easily solvable.

We denote the discrete space of the domain $\Omega$ in the CFD simulation by $\Omega_D = \{\mathbf{y}_1, \mathbf{y}_2,...,\mathbf{y}_N\}$, where $N$ is the number of mesh points. We assume that $\Omega_s\subset \Omega_D$, which implies that each sensing location is constrained to be one of the mesh points. Hence, we can denote $\mathbf{x}_k = \mathbf{y}_{l_k}$ and $\Omega_s=\{\mathbf{y}_{l_1}, \mathbf{y}_{l_2},... \mathbf{y}_{l_m}\}$, where $l_k$ is the mesh index of the $k$-th sensing location. Then, we can approximate the cost function $J_0$ in \eqref{eq:EnergyFun} as
\begin{equation}\label{eq:Japprox}
  J_0\approx J_1 = \sum_{i=1}^{N}\sum_{k=1}^{m} w_{il_k}(v_i-s_k)^2 +\lambda\sum_{i}^{N}\sum_{j}^{N}w_{ij}(v_i-v_j)^2
\end{equation}
where $w_{ij}=w(\mathbf{y}_i, \mathbf{y}_j)$ and $v_i = v(\mathbf{y}_i)$.
We introduce a new matrix $\mathbf{W}=[w_{ij}]$, i.e., $w_{ij}$ is the entry of $\mathbf{W}$ located at the $i$-th row and $j$-th column.

For a more elegant description,we use the following cost function $J$ instead of $J_1$
\begin{eqnarray}
  J&=&\frac{J_1}{\lambda}  = \frac{1}{\lambda}\sum_{i=1}^{N}\sum_{k=1}^{m} w_{il_k}(v_i-s_k)^2 + \sum_{i}^{N}
                              \sum_{j}^{N}w_{ij}(v_i-v_j)^2 \nonumber \\
                         &=& \frac{1}{\lambda}\sum_{k=1}^{m}(\mathbf{v}-s_k\mathbf{1})^\mathrm{T}W_{l_k}^\mathrm{D}
                              (\mathbf{v}-s_k\mathbf{1}) + \sum_{j=1}^{N}\mathbf{v}^\mathrm{T}(I-A_j)^\mathrm{T}W_j^\mathrm{D}(I-A_j)\mathbf{v} \nonumber \\
                         &=& \mathbf{v}^\mathrm{T}\left(\frac{1}{\lambda}\sum_{k=1}^{m}W_{l_k}^\mathrm{D}+ \sum_{j=1}^{N}(I-A_j)^\mathrm{T}W_j^\mathrm{D}(I-A_j)\right)\mathbf{v} -\left(\frac{2}{\lambda}\sum_{k=1}^{m}s_k\mathbf{w}_{l_k}^\mathrm{T}\right)\mathbf{v} + \frac{1}{\lambda}\sum_{k=1}^{m}s_k^2\mathrm{Tr}(W_{l_k}^\mathrm{D})\nonumber \\
                         &=& \mathbf{v}^\mathrm{T}H\mathbf{v}-2\mathbf{b}^\mathrm{T}\mathbf{v}+\mathbf{c} \label{eq:J}
\end{eqnarray}
where $\mathbf{v}=[v_1,v_2,..., v_N]^\mathrm{T}$, $\mathbf{1}\in\mathbb{R}^N$ is a vector with all 1 entries, $W_j^\mathrm{D}=\mathrm{diag}\{\mathbf{w}_j\}$, $\mathbf{w}_j$ is the $j$-th column of $\mathbf{W}$, and $A_j$ is a matrix with all 1 entries in $j$-th column and all 0 entries in others. Here, $\mathrm{diag}\{\cdot\}$ and $\mathrm{Tr}(\cdot)$ are diagonal matrix and trace operators, respectively. It is clear that
\begin{equation*}
  \mathbf{b} = \frac{1}{\lambda}\sum_{k=1}^{m}s_k\mathbf{w}_{l_k}
\end{equation*}
and
\begin{eqnarray}
  H &=& \frac{1}{\lambda}\sum_{k=1}^{m}W_{l_k}^\mathrm{D}+ \sum_{j=1}^{N}(I-A_j)^\mathrm{T}W_j^\mathrm{D}(I-A_j) \nonumber \\
   &=& \frac{1}{\lambda}\sum_{k=1}^{m}W_{l_k}^\mathrm{D}+ \sum_{j=1}^{N} W_j^\mathrm{D}- A_j^\mathrm{T}W_j^\mathrm{D} - W_j^\mathrm{D}A_j + A_j^\mathrm{T}W_j^\mathrm{D}A_j \nonumber \\
   &=& \frac{1}{\lambda}\sum_{k=1}^{m}W_{l_k}^\mathrm{D} +\mathrm{diag} \left\{ 2\sum_{j=1}^{N}\mathbf{w}_j \right\} -2\mathbf{W} \nonumber \\
   &=&D-W_s \label{eq:H}
\end{eqnarray}
where $\sum_{j=1}^{N}W_j^\mathrm{D}=\mathrm{diag}\{\sum_{j=1}^{N}\mathbf{w}_j\}$, $\sum_{j=1}^{N}A_j^\mathrm{T}W_j^\mathrm{D}=\mathbf{W}^\mathrm{T}=\mathbf{W}$, $\sum_{j=1}^{N}W_j^\mathrm{D}A_j^\mathrm{T}=\mathbf{W}$, and $A_j^\mathrm{T}W_j^\mathrm{D}A_j=\mathrm{diag}\{\sum_{j=1}^{N}(\mathbf{w}_j^\mathrm{r})^\mathrm{T}\}=\mathrm{diag}\{\sum_{j=1}^{N}\mathbf{w}_j\}$. Here, $\mathbf{w}_j^\mathrm{r}$ is the $j$-th row of $\mathbf{W}$. Since $\mathbf{W}$ is symmetric, $(\mathbf{w}_j^\mathrm{r})^\mathrm{T}=\mathbf{w}_j$.

The matrix $D$ in \eqref{eq:H} is a diagonal matrix, and $D=\frac{1}{\lambda}\sum_{k=1}^{m}W_{l_k}^\mathrm{D} +\mathrm{diag} \left\{ 2\sum_{j=1}^{N}\mathbf{w}_j \right\}$. It is clear that $W_s = 2\mathbf{W}$.

To minimize the cost function $J$, we take the derivative of \eqref{eq:J} with respective to $\mathbf{v}$ and set it as zero, i.e.,
\begin{equation}\label{eq:DiffJ}
  \frac{\mathrm{d}J}{\mathrm{d}\mathbf{v}} = 2(H\mathbf{v}-\mathbf{b})=0
\end{equation}
From \eqref{eq:DiffJ}, we can easily obtain
\begin{equation}\label{eq:solutionv}
  \hat{\mathbf{v}}=H^{-1}\mathbf{b}=(D-W_s)^{-1}\mathbf{b}
\end{equation}

Substituting $\hat{\mathbf{v}}$ into \eqref{eq:cali}, we obtain the calibrated physical field \begin{equation}\label{eq:fcalibrated}
 {f}(\mathbf{y}_i)=f_c(\mathbf{y}_i)-\hat{v}_i, \quad \mathbf{y}_i\in\Omega_\mathrm{D}\; \mathrm{and} \;1\leq i\leq N
\end{equation}

For the matrix $H\in \mathbb{R}^{N\times N}$, $N$ is the number of mesh points. Generally, $N$ is very large and therefore, the computational cost of solving $H^{-1}$ in \eqref{eq:solutionv} is extremely expensive. In practice, we cannot directly obtain $\hat{\mathbf{v}}$ from \eqref{eq:solutionv}. Thus, we provide a computationally efficient method to find the approximated solution of $\hat{\mathbf{v}}$.

\subsection{Computational efficient approximated solution }
To avoid directly solving the inverse of the matrix $H$, which has very large dimension, we provide an approximated solution of the inverse problem. A similar method has been previously applied for image editing\cite{an2008appprop}.


Considering the definition of $w_{ij}$, i.e., $w_{ij}=w(\mathbf{y}_i,\mathbf{y}_j)$, we find that each column of $\mathbf{W}$ represents the affinities between the point ( whose mesh index is the column index) and the whole physical field. The neighbouring points have similar affinities with other points. Hence, the neighbouring columns have similar entries, especially for the entries far away from the diagonal, which implies that the neighbouring columns of $\mathbf{W}$ have a large correlation coefficient. Therefore, the matrix $\mathbf{W}$ has many near zero eigenvalues, and it can be approximated by a low rank matrix.

Since the matrix $W_s=2\mathbf{W}\in \mathbb{R}^{N\times N}$, $W_s$ can also be approximated by a low rank matrix. We rewrite the symmetric matrix $W_s$ as
\begin{equation}\label{eq:Ws}
  W_s = \left[\begin{array}{cc}
                A & B^T \\
                B & C
              \end{array} \right]
\end{equation}
where $A\in\mathbb{R}^{n\times n}$, $B\in \mathbb{R}^{(N-n)\times n}$, and $C\in \mathbb{R}^{(N-n)\times(N-n)}$.

\begin{figure}[htb]
    \centering
        \includegraphics[width=0.6\textwidth]{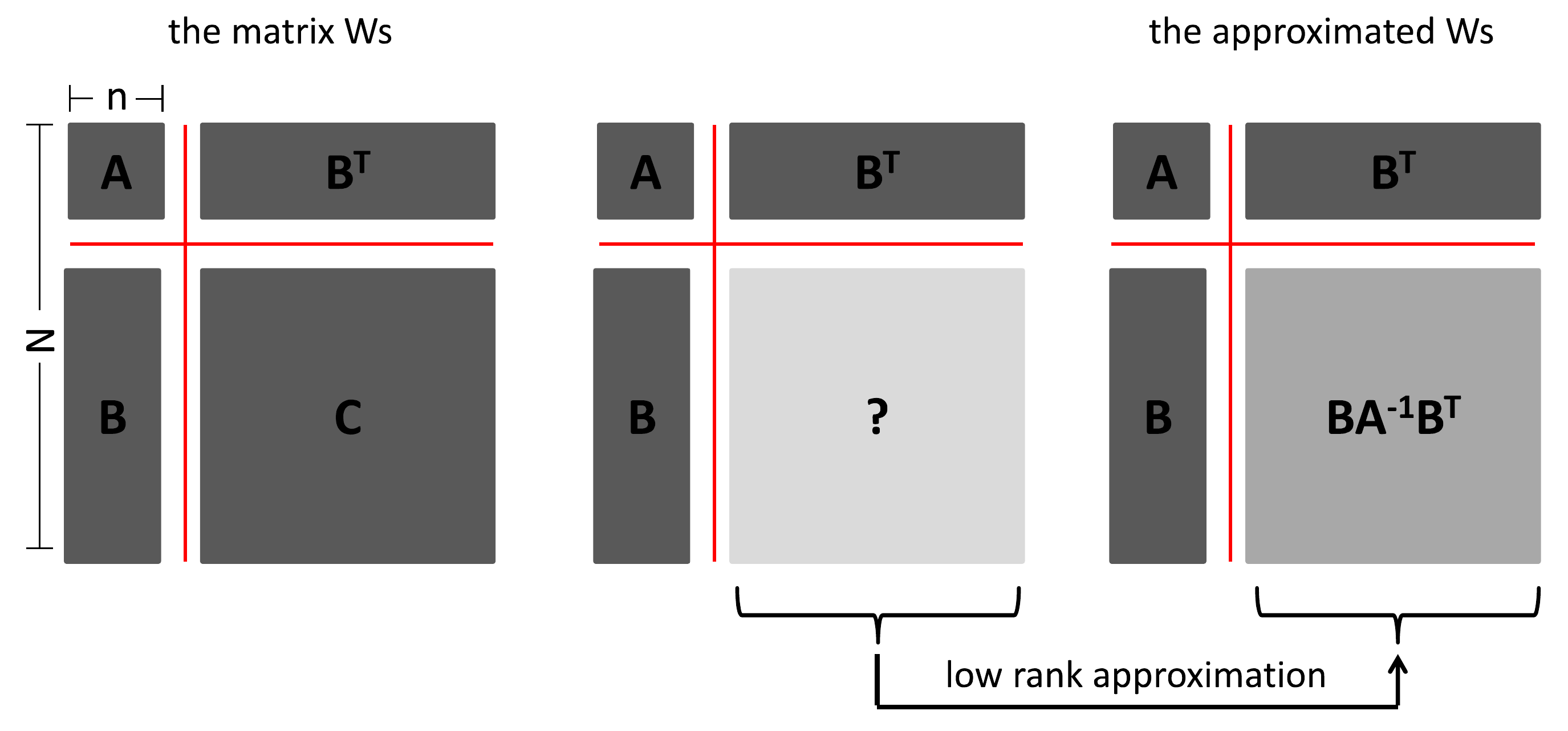}
 \caption{Illustration of the low rank approximation of $W_s$. (This figure was firstly used in \cite{an2008appprop}.)}
    \label{Fig:LowRankApprox}
\end{figure}

We assume that the number of dominant eigenvalues of $W_s$ is $n$. Using the eigen decomposition and ignoring all the near zero eigenvalues, we can approximate $W_s$ as
\begin{equation*}
  W_s =U\Lambda U^\mathrm{T}\approx U\left[\begin{array}{cc}
                 \Lambda_1 & \mathbf{0} \\
                 \mathbf{0} & \mathbf{0}
               \end{array}
  \right]U^\mathrm{T}=VV^\mathrm{T}
\end{equation*}
where $U$ is an orthogonal matrix, $\Lambda$ is a diagonal matrix with all eigenvalues of $W_s$ in the diagonal, $\Lambda_1=\Lambda(1:n, 1:n)$ consists of all dominant eigenvalues of $W_s$ in its diagonal, and $V\in \mathbb{R}^{N\times n}$ has exactly rank $n$. We denote $V$ by
\begin{equation*}
  V=\left[\begin{array}{cc}
            X \\
            Y
          \end{array}
  \right]
\end{equation*}
where $X$ and $Y$ have the same sizes as $A$ and $B$ in \eqref{eq:Ws}, respectively. Then, we can obtain that
\begin{equation}\label{eq:WsV}
  W_s\approx  \left[\begin{array}{cc}
            X \\
            Y
          \end{array}\right] \left[\begin{array}{cc}
                                     X^\mathrm{T} & Y^\mathrm{T}
                                   \end{array} \right]
                                   =\left[\begin{array}{cc}
                                            XX^\mathrm{T} & XY^\mathrm{T} \\
                                            YX^\mathrm{T} & YY^\mathrm{T }
                                          \end{array}
                                   \right]
\end{equation}
Comparing \eqref{eq:WsV} and \eqref{eq:Ws}, as illustrated in Fig. \ref{Fig:LowRankApprox},  we can find
\begin{equation}\label{eq:C}
  C\approx YY^\mathrm{T}=YX^\mathrm{T} ((X^\mathrm{T})^{-1}X^{-1})XY^\mathrm{T}=BA^{-1}B^\mathrm{T}
\end{equation}

Substituting \eqref{eq:C} into \eqref{eq:Ws} yields
\begin{equation}\label{eq:WsApprox}
  W_s\approx \left[\begin{array}{cc}
                     A & B^\mathrm{T} \\
                     B & BA^{-1}B^\mathrm{T}
                   \end{array} \right] = \left[\begin{array}{c}
                                                 A \\
                                                 B
                                               \end{array}
                   \right]A^{-1}\left[\begin{array}{cc}
                                      A & B^\mathrm{T}
                                    \end{array}
                   \right]=ZA^{-1}Z^\mathrm{T}
\end{equation}
where $Z=\left[\begin{array}{c}
            A \\
            B
          \end{array}\right]$
is the first $n$ column of $W_s$. Using this way, we can find a low rank approximation of $W_s$, which can both resolve the storage and computational cost problems in solving $H^{-1}$.

Substituting the approximation of $W_s$ in \eqref{eq:WsApprox} into \eqref{eq:H} and using the Woodbury formula \cite{golub2012matrix}, we can obtain
\begin{equation}\label{eq:Hinv}
  H^{-1} \approx (D-ZA^{-1}Z^\mathrm{T})^{-1}  = D^{-1}-D^{-1}Z(-A+Z^TD^{-1}Z)^{-1}Z^{T}D^{-1}
\end{equation}

Then, substituting \eqref{eq:Hinv} into \eqref{eq:solutionv} yields
\begin{equation}\label{eq:v}
  \hat{\mathbf{v}}=D^{-1}\mathbf{b}-D^{-1}Z(-A+Z^\mathrm{T}D^{-1}Z)^{-1}Z^{\mathrm{T}}D^{-1}\mathbf{b}
\end{equation}
In \eqref{eq:v}, we need to solve the inverse of two matrices. $D$ is a diagonal matrix and its inverse can be easily solved. Because $(-A+Z^\mathrm{T}D^{-1}Z)\in \mathbb{R}^{n\times n}$ and $n\ll N$, solving $(-A+Z^\mathrm{T}D^{-1}Z)^{-1}$ is computationally much more efficient.

Substituting \eqref{eq:v} into \eqref{eq:fcalibrated}, we can obtain the calibrated physical field. Next, we show an experiment to support the proposed methodology.

\section{Experimental verification}

\subsection{Statement of the experiment}
\begin{figure}[htb]
    \centering
    \begin{subfigure}[b]{0.45\textwidth}
        \centering
        \includegraphics[width=\textwidth]{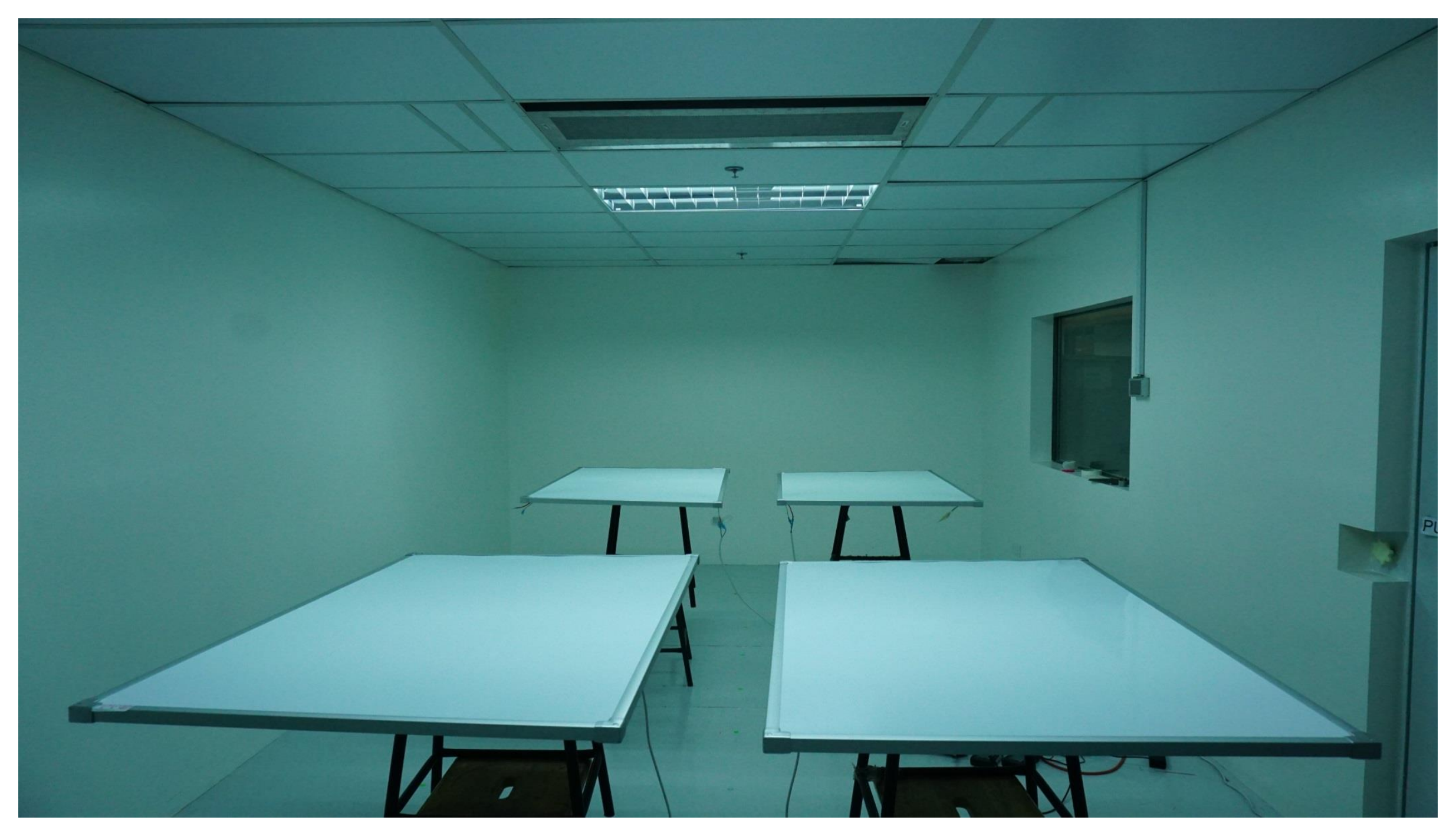}
    \end{subfigure}%

    \begin{subfigure}[b]{0.45\textwidth}
        \centering
        \includegraphics[width=\textwidth]{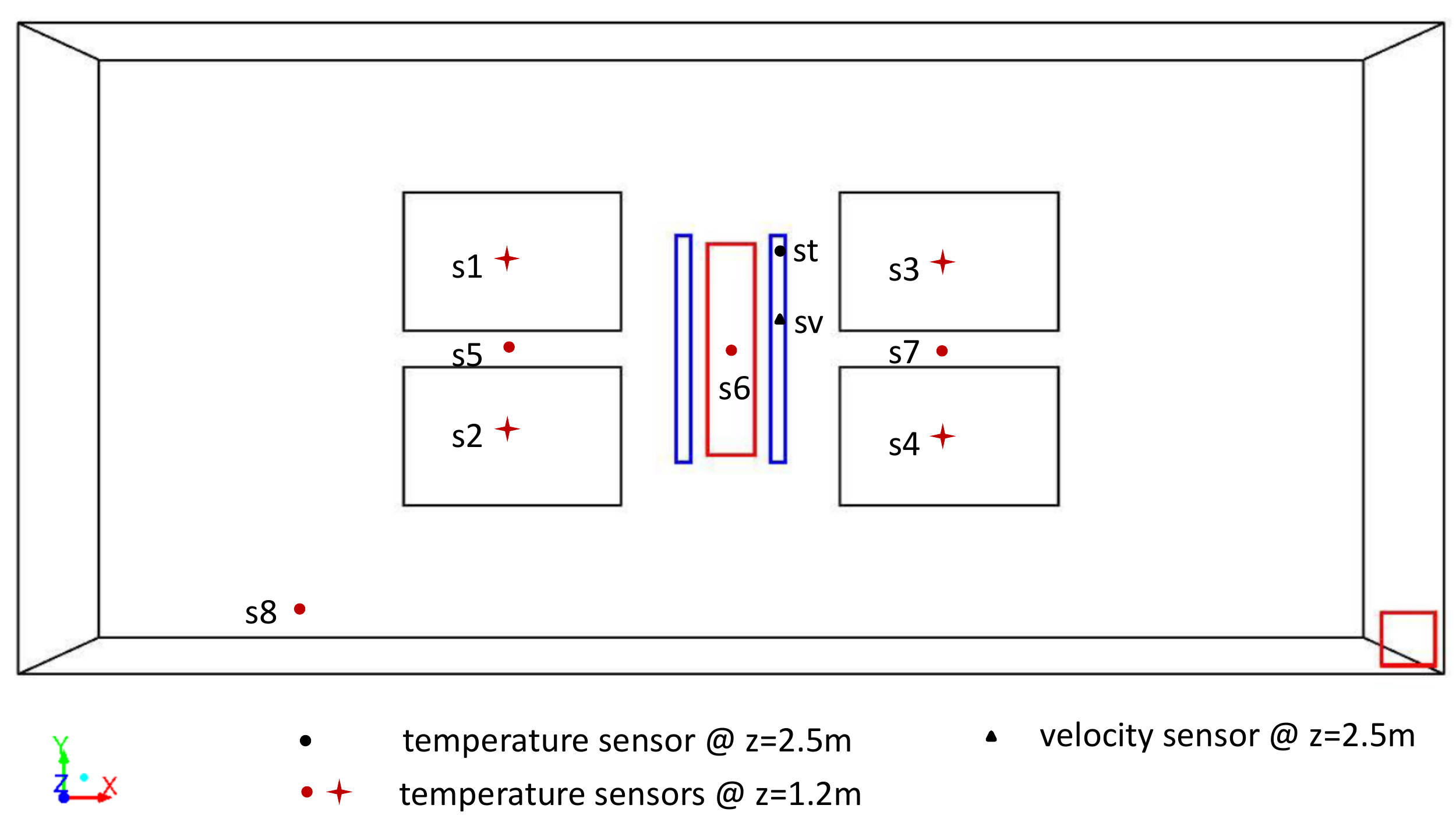}
    \end{subfigure}
 \caption{The test bed: a $7.28\mathrm{m} \times 3.32\mathrm{m}\times 2.5\mathrm{m}$ room located at the Process Control and Instrumentation Lab in Nanyang Technological University, Singapore. The bottom figure is the top view sketch map. The red square and the red rectangle are the two outlets. The center one is for induced air, and the corner one is for exhaust air. The two blue rectangles represent the two cooling air inlets. The black triangle represents an air velocity sensor, as shown in the left-most picture of Fig. \ref{Fig:Sensors}. The black dot represents a temperature sensor as shown in the middle picture of Fig. \ref{Fig:Sensors}. The four red stars and four red dots represent eight temperature sensors located 1.2m from the ground. The four black squares are four heaters. The height of the heaters were 1m above the ground.}
    \label{Fig:Room}
\end{figure}

We tested the proposed methodology for a 7.28m$\times$3.32m$\times$2.5m room, which is shown in Fig. \ref{Fig:Room}.  In the room, an Active Chilled Beam system (ACBs) was installed for air conditioning, as shown on the ceiling of the top picture in Fig. \ref{Fig:Room}. One air outlet is at the corner of the room. The walls and the door are made by thermal insulation materials, and the windows are double glazing. Hence, in the CFD simulation, we simply set the walls, the door, and the windows as insulated walls.

\begin{figure}[htb]
    \centering
    \begin{subfigure}[b]{0.16\textwidth}
        \centering
        \includegraphics[width=\textwidth]{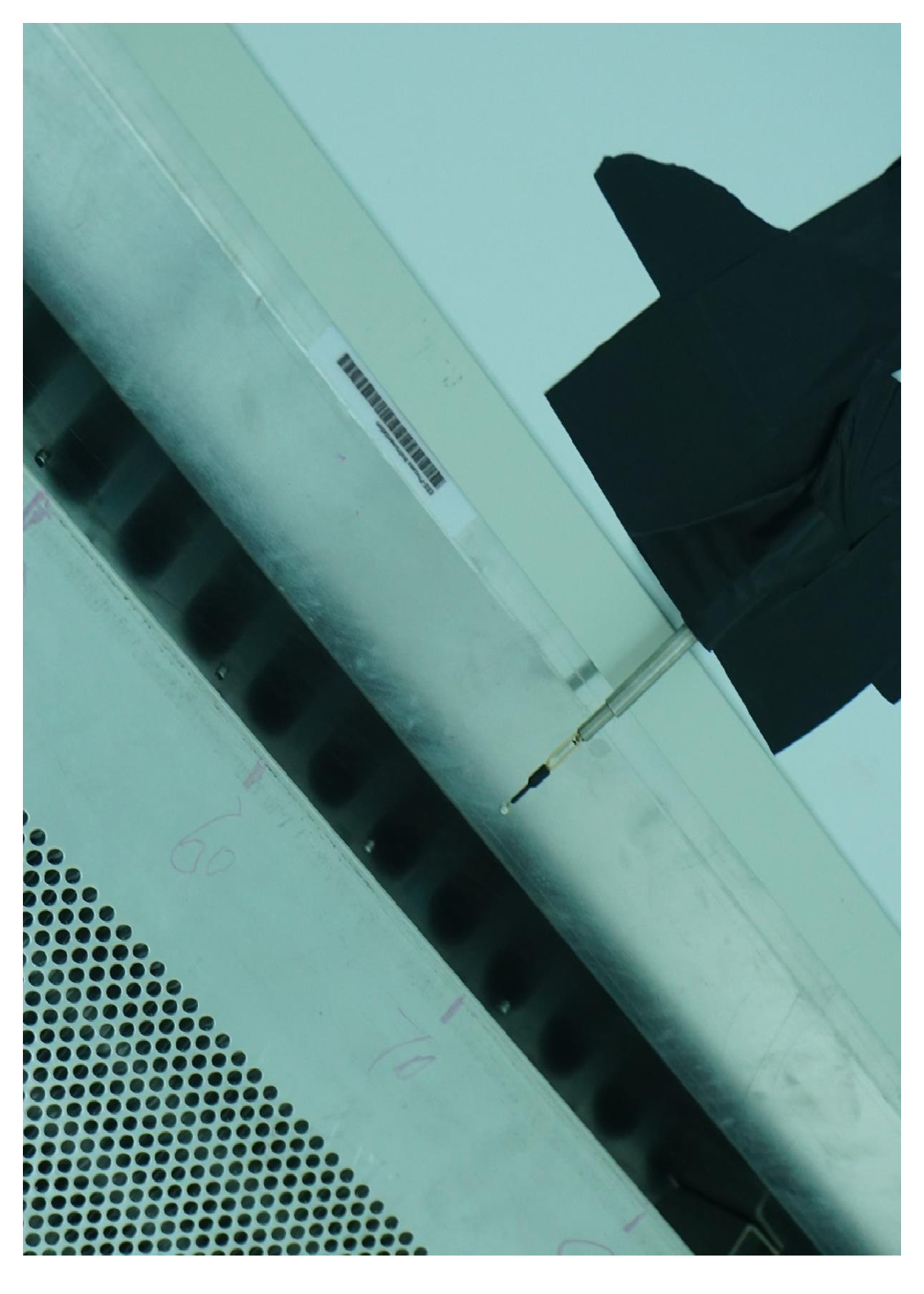}
    \end{subfigure}%
    \begin{subfigure}[b]{0.16\textwidth}
    \centering
    \includegraphics[width=\textwidth]{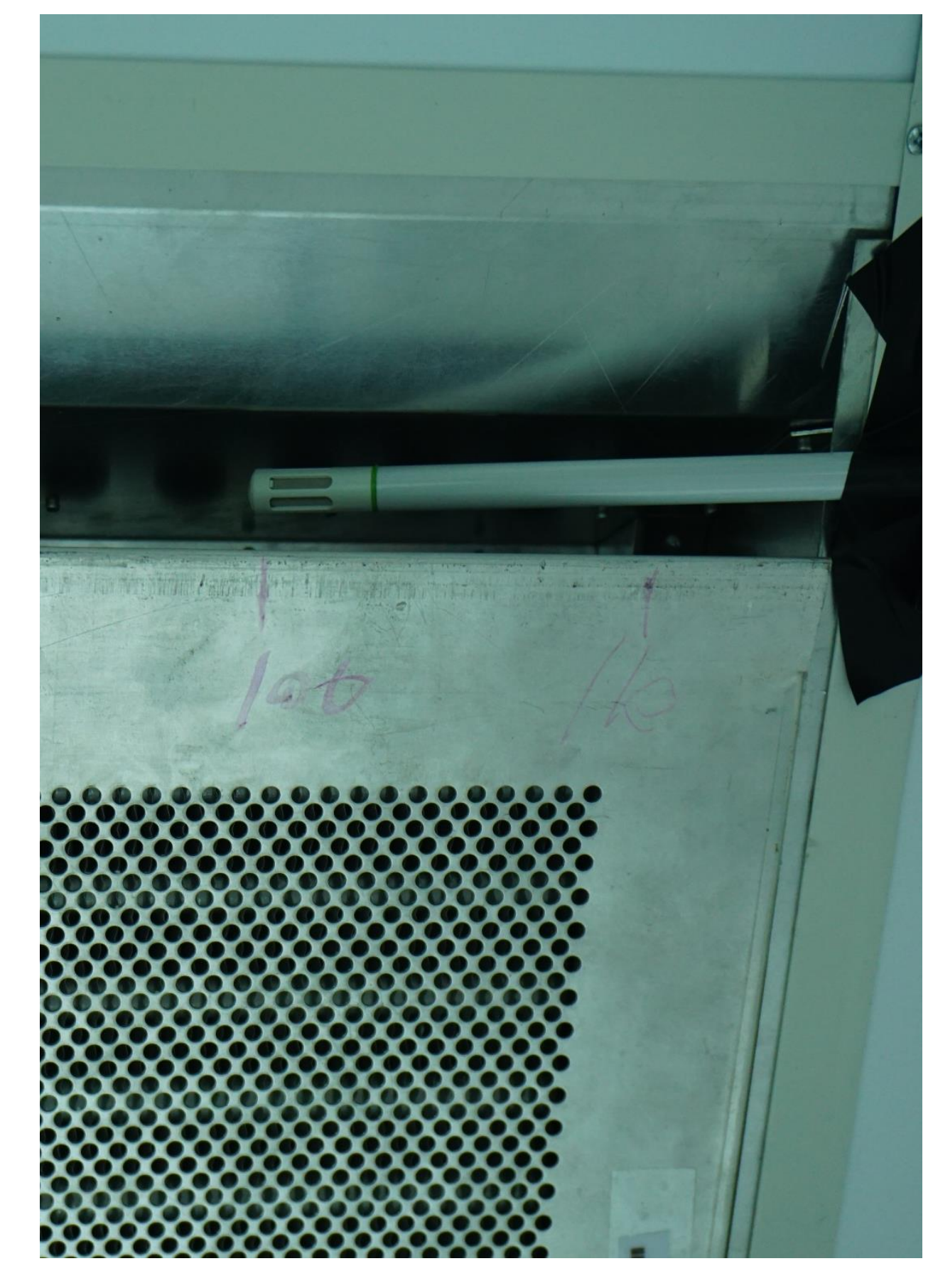}
\end{subfigure}
    \begin{subfigure}[b]{0.16\textwidth}
        \centering
        \includegraphics[width=\textwidth]{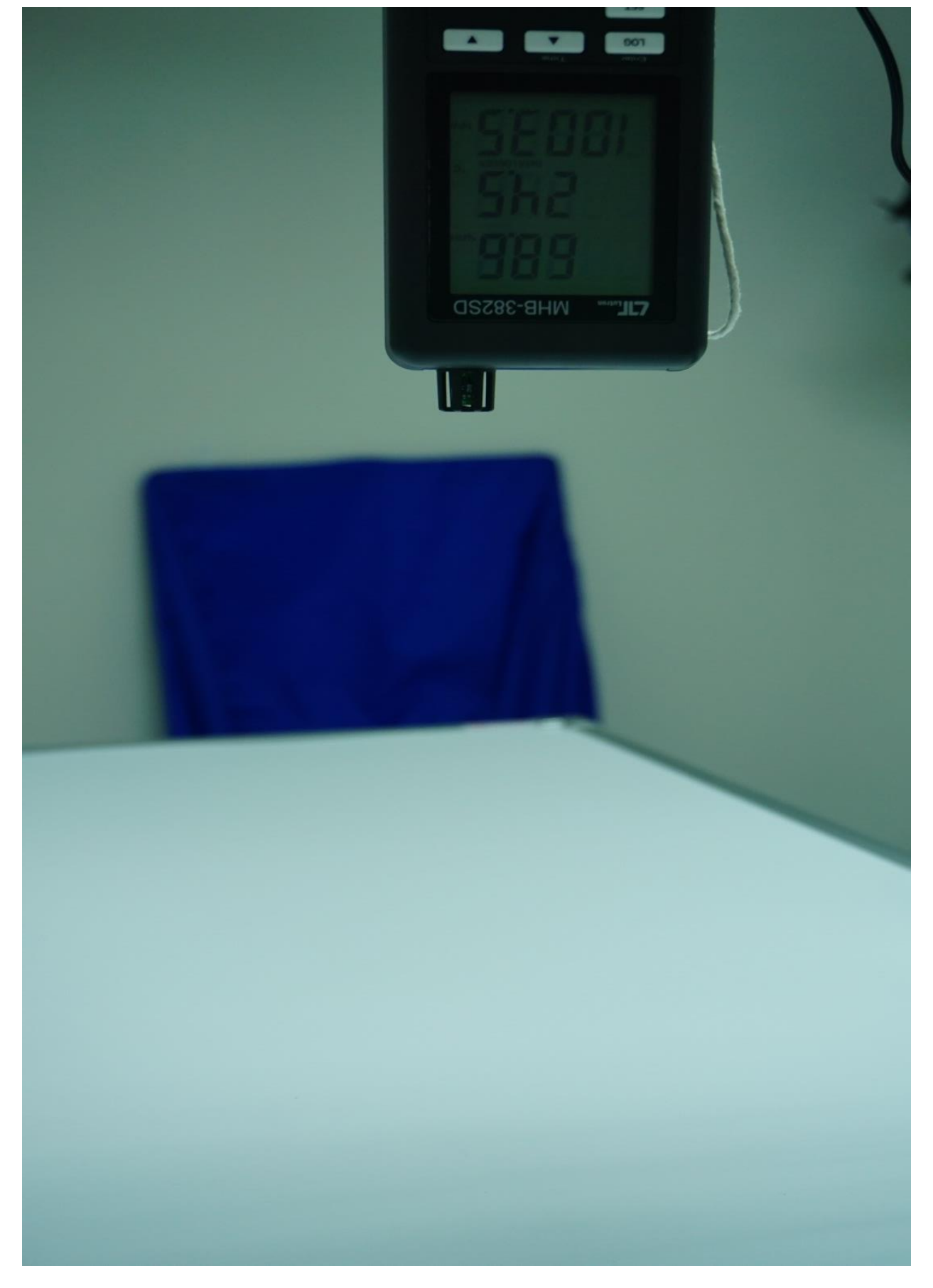}
    \end{subfigure}

 \caption{Sensors used for indoor parameters measurement. Left: velocity sensor used to measure the inlet air velocity. Middle: temperature sensor used to measure the inlet cooling air temperature. Right: temperature sensor used to measure the temperature at 1.2m height, which is hung above the heater. }
    \label{Fig:Sensors}
\end{figure}

As shown in Fig. \ref{Fig:Room}, the room contains four heaters (0.76m$\times$1.2m$\times$0.02m). As shown in the bottom picture of Fig. \ref{Fig:Room}, the left two heaters are 300W, and the other two are 400W. The height above the ground of the top surfaces of the heaters is $1.02\mathrm{m}$. The bottom faces of the heaters were made by insulated materials. One sensor was used to measure the velocity of the inlet cooling air, as shown in the left-most picture of Fig. \ref{Fig:Sensors}. Another sensor was used to measure the inlet cooling air temperature, which is shown in the middle picture in Fig. \ref{Fig:Sensors}. The other eight sensors were used to measure the temperature of a horizontal plane 1.2m above the ground. Four sensors were hung and suspended just above centers of the four heaters, respectively. The four sensors are represented by stars (s1-s4) in the bottom picture of Fig. \ref{Fig:Room}. One of the four is shown in the right-most picture of Fig. \ref{Fig:Sensors}. Three sensors (s5-s7) were placed at the middle plane of the room. The sensor s8 was placed 20cm from the wall. All the eight sensors (s1-s8) were 1.2m above the ground.

\begin{figure}[htb]
    \centering
    \begin{subfigure}[b]{0.75\textwidth}
        \centering
        \includegraphics[width=\textwidth]{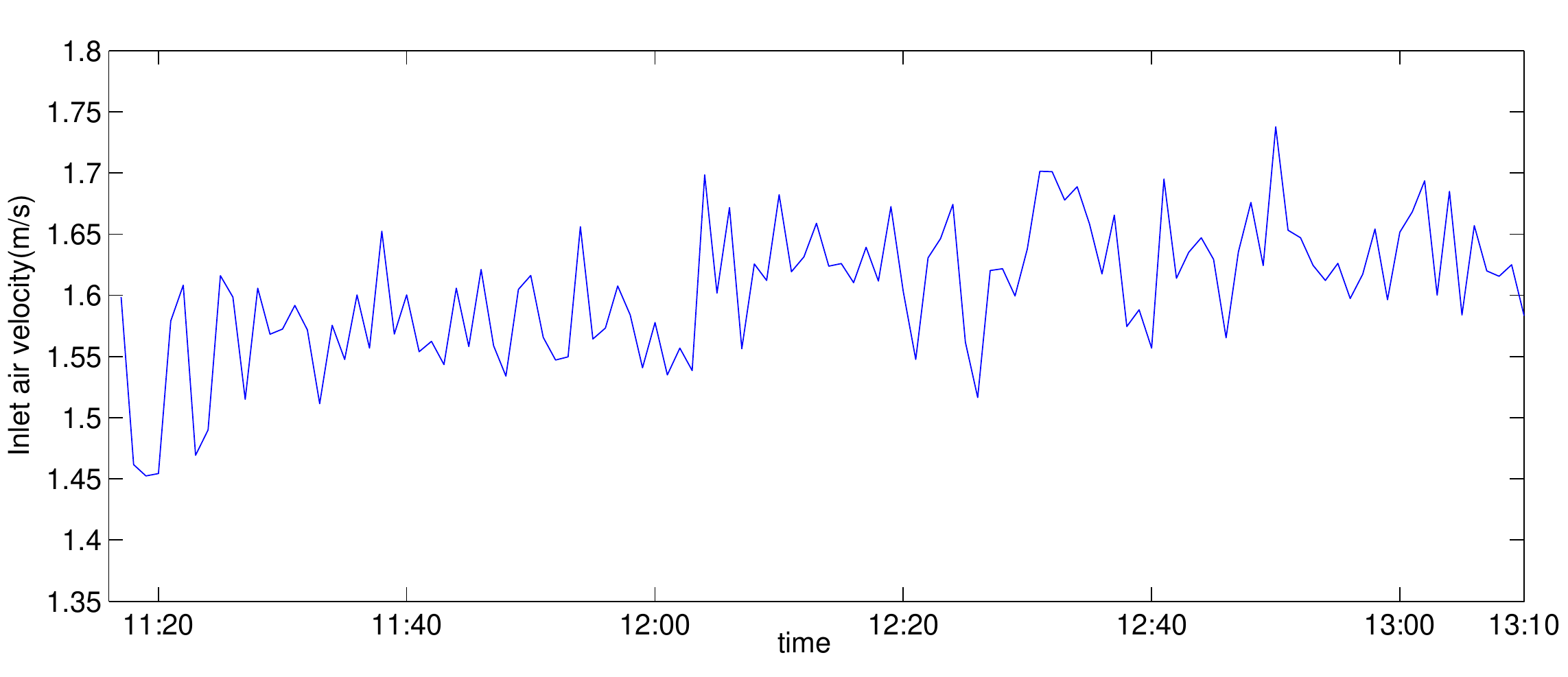}
    \end{subfigure}%

    \begin{subfigure}[b]{0.75\textwidth}
        \centering
        \includegraphics[width=\textwidth]{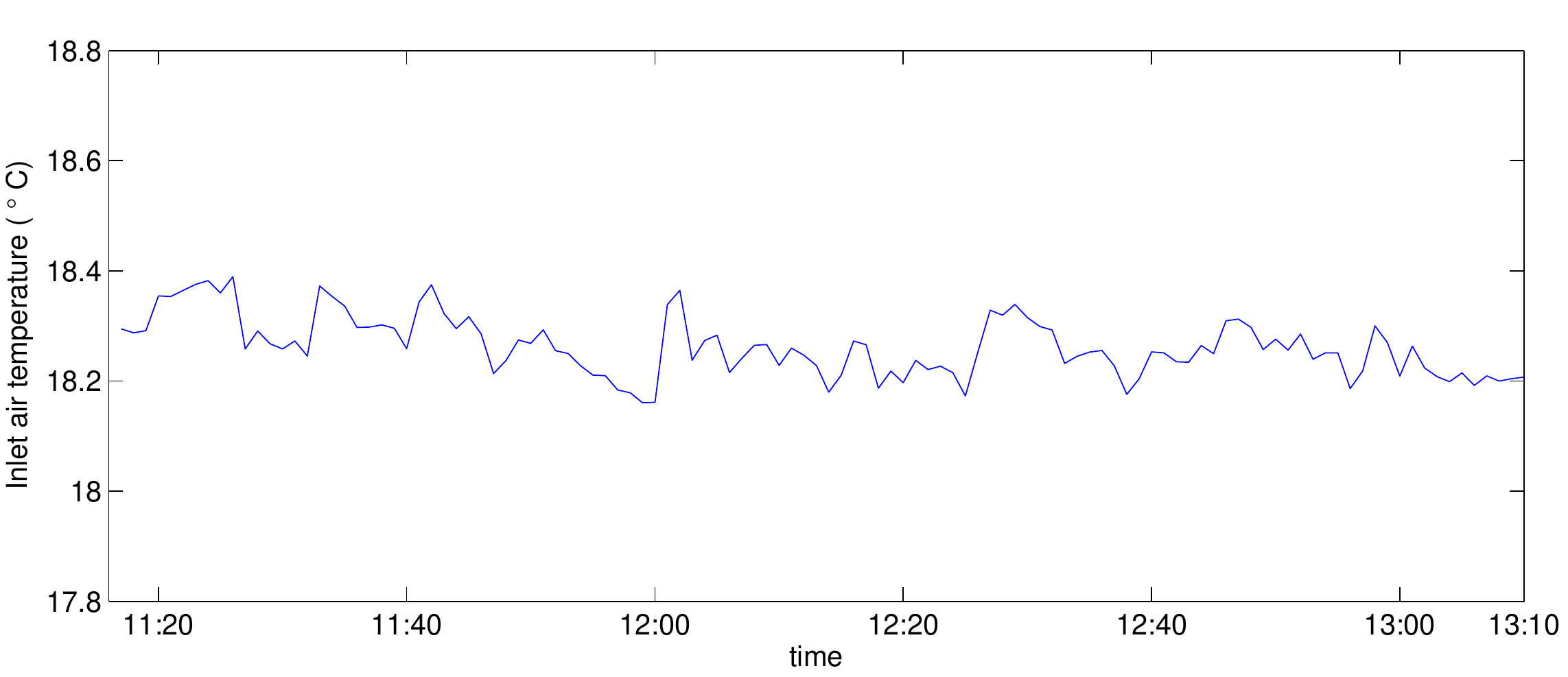}
    \end{subfigure}
 \caption{The recorded inlet cooling air velocity and temperature during 11:17 to 13:10 on 26th May, 2016.}
    \label{Fig:InletData}
\end{figure}

\begin{figure}[htb]
    \centering
        \includegraphics[width=0.75\textwidth]{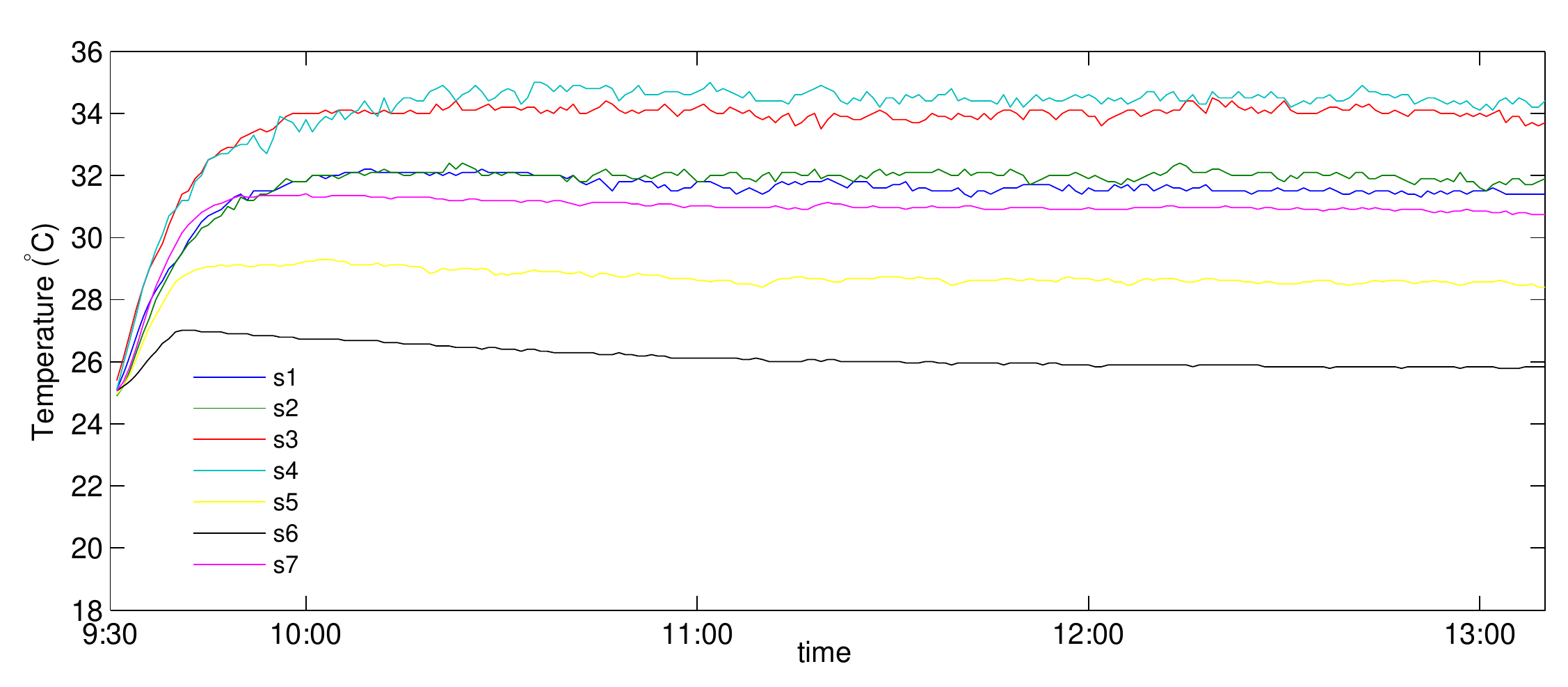}
 \caption{The temperature recording of the seven sensors (s1-s7).}
    \label{Fig:RawTempData}
\end{figure}

The experiment was carried out on 26th May, 2016. The four heaters and the air-condition system were turned on at 9:30 am. We collected the velocity and temperature of the inlet cooling air (via the sensors $\mathrm{s}_v$ and $\mathrm{s}_t$, respectively) from 11:17 to 13:10. The data is shown in Fig. \ref{Fig:InletData}. The sensors s1-s7 recorded the temperature of the 1.2m plane from 9:30 to 13:10, the data for which is shown in Fig. \ref{Fig:RawTempData}. The sensor s8 does not store the sensor readings. We have recorded that after 11:00, the readings of sensor s8 remained at $25.6^\circ C$.

\subsection{CFD simulation}
The simulation was performed using the commercial CFD software Ansys FLUENT v.13.0. The proposed calibration methodology has the potential to be applied for dynamic CFD simulations if both the real-time simulation results and sensor observations are available. In this work, however, for simplicity we only consider the steady state simulation results.

The indoor air was assumed as the ideal gas with density of $1.225\; \mathrm{kg/m}^3$. The viscous model was chosen as the realizable $k-\varepsilon$ model. Since the test-bed was built in a laboratory, we did not consider the solar radiation. The door, ground, ceiling, windows, and walls are assumed to be insulated walls. During the experiment, all lights were turned off. For the velocity and temperature of the inlet cooling air, we use the mean values of the sensor readings in the time period 12:10 to 13:10 in Fig. \ref{Fig:InletData}. The bottom faces of the heaters were made by insulated material, but they were not completely insulated. We assumed that $5\%$ of energy was transferred via the bottom faces of the heaters. Corresponding to the four sensors s1-s4, we denote the four heaters by heater 1, heater 2, heater 3 and heater 4, respectively. The boundary conditions of the CFD simulation are detailed in Table \ref{Table:BC}.

\begin{table*}[tbh]
\caption{Boundary conditions of the CFD simulation}
\begin{center}
\begin{tabular}{llll}
\hline
Boundary & Type & Heat transfer & Mass \& momentum\\
\hline
The two inlets & velocity-inlet & $T_i=18.24 ^\circ \mathrm{C}$ & $V_{in}=1.62 \mathrm{m/s}, \theta = 25^\circ$ \\
Center outlet & pressure-outlet & $T_o = 25 ^\circ \mathrm{C}$ & $P_{\mathrm{relative}}=-10\mathrm{Pa}$ \\
Corner outlet & pressure-outlet & $T_o = 25 ^\circ\mathrm{ C}$ & $P_{\mathrm{relative}}=0\mathrm{Pa}$ \\
Walls, windows, ceiling, \& ground & wall & Insulated & No slip wall\\
Top and side faces of heaters 1\&2& wall & $h_c = 287.8 \mathrm{W/m^2}$ & No slip wall\\
Bottom faces of heaters 1\&2 & wall & $h_c = 16.4 \mathrm{W/m^2}$ & No slip wall\\
Top and side faces of heaters 3\&4& wall & $h_c = 383.7 \mathrm{W/m^2}$ & No slip wall\\
Bottom faces of heaters 3\&4 & wall & $h_c = 21.9 \mathrm{W/m^2}$ & No slip wall\\
\hline\hline
\end{tabular}
\begin{tablenotes}
        \footnotesize
$T_i-$ inlet cooling air temperature; $V_\mathrm{in}-$ inlet cooling air speed; $\theta-$ the angle of $V_\mathrm{in}$ with the ceiling; $T_o-$ outlet air temperature; $P_{\mathrm{relative}}-$ pressure relative to the reference pressure ($P_{\mathrm{reference}}=101325\mathrm{Pa}$); $h_c-$ convective heat flux.
\end{tablenotes}
\label{Table:BC}
\end{center}
\end{table*}

We created the structured mesh using Gambit v2.4. Since all the walls are insulated, we did not consider the boundary layer of the walls and uniformly divided the horizontal plane ($X-Y$ plane) into $2\mathrm{cm}\times 2\mathrm{cm}$ squares. In the $z$-direction, we made the mesh dense near the ceiling ($z=2.5m$) and heater top faces ($z = 1.02m$), and divided the entire 2.5m height into 137 intervals. The total mesh size is $\Delta_1=167\times365\times137-4\times37\times59\times1=8,342,103$.

To verify the grid independency, we made the mesh dense and increased the mesh size to $\Delta_2=18,527,205$. 
We did CFD simulations for both mesh files. The results were found to be in good agreement. Therefore, the mesh file with 8,342,103 cells is proper and the number of cells is large enough for our experiment.

\begin{figure}[htb]
    \centering
        \includegraphics[width=0.65\textwidth]{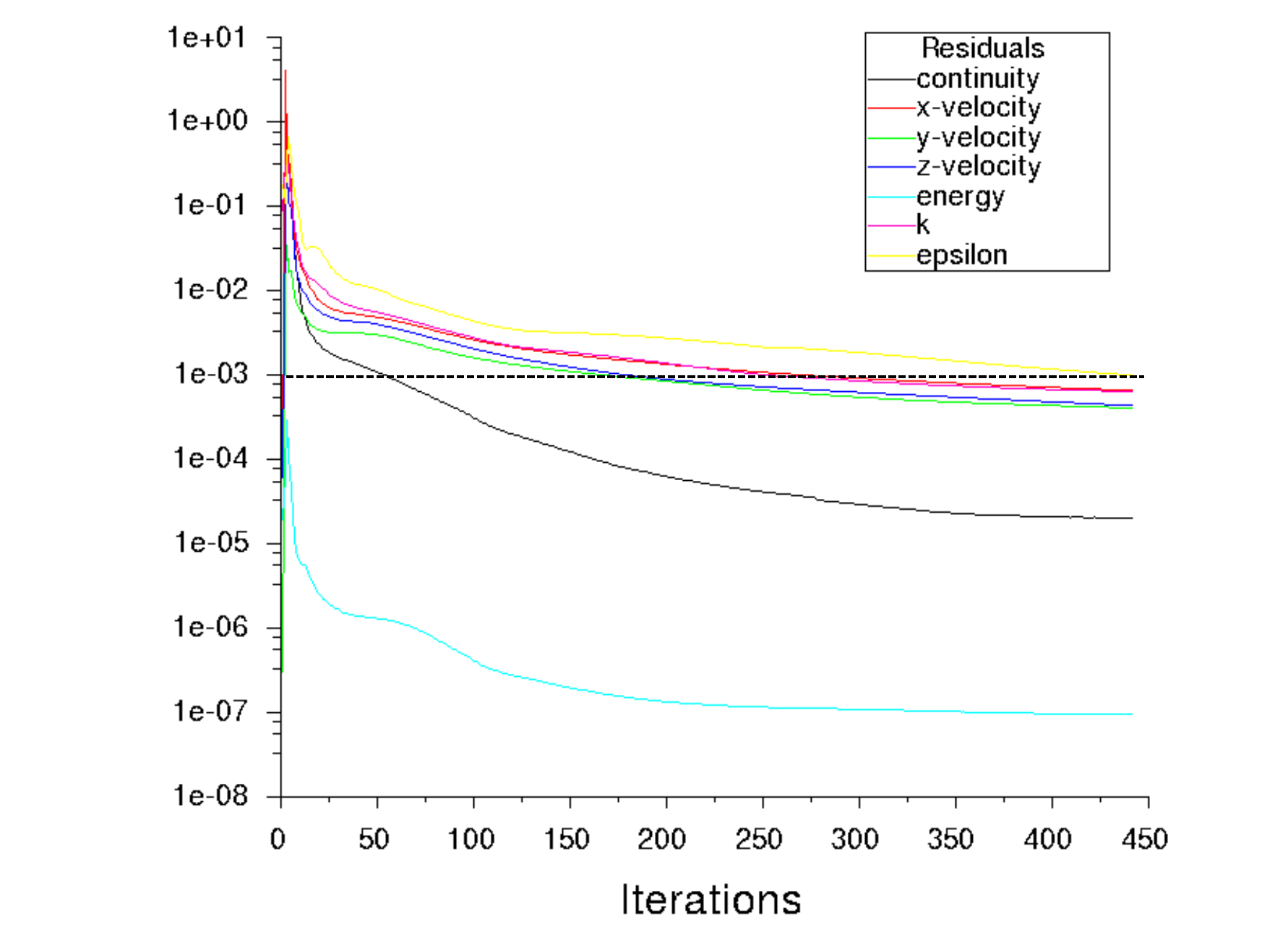}
 \caption{Convergence history of the residuals.}
    \label{Fig:converge}
\end{figure}

\begin{figure}[htb]
    \centering
    \begin{subfigure}[b]{0.75\textwidth}
        \centering
        \includegraphics[width=\textwidth]{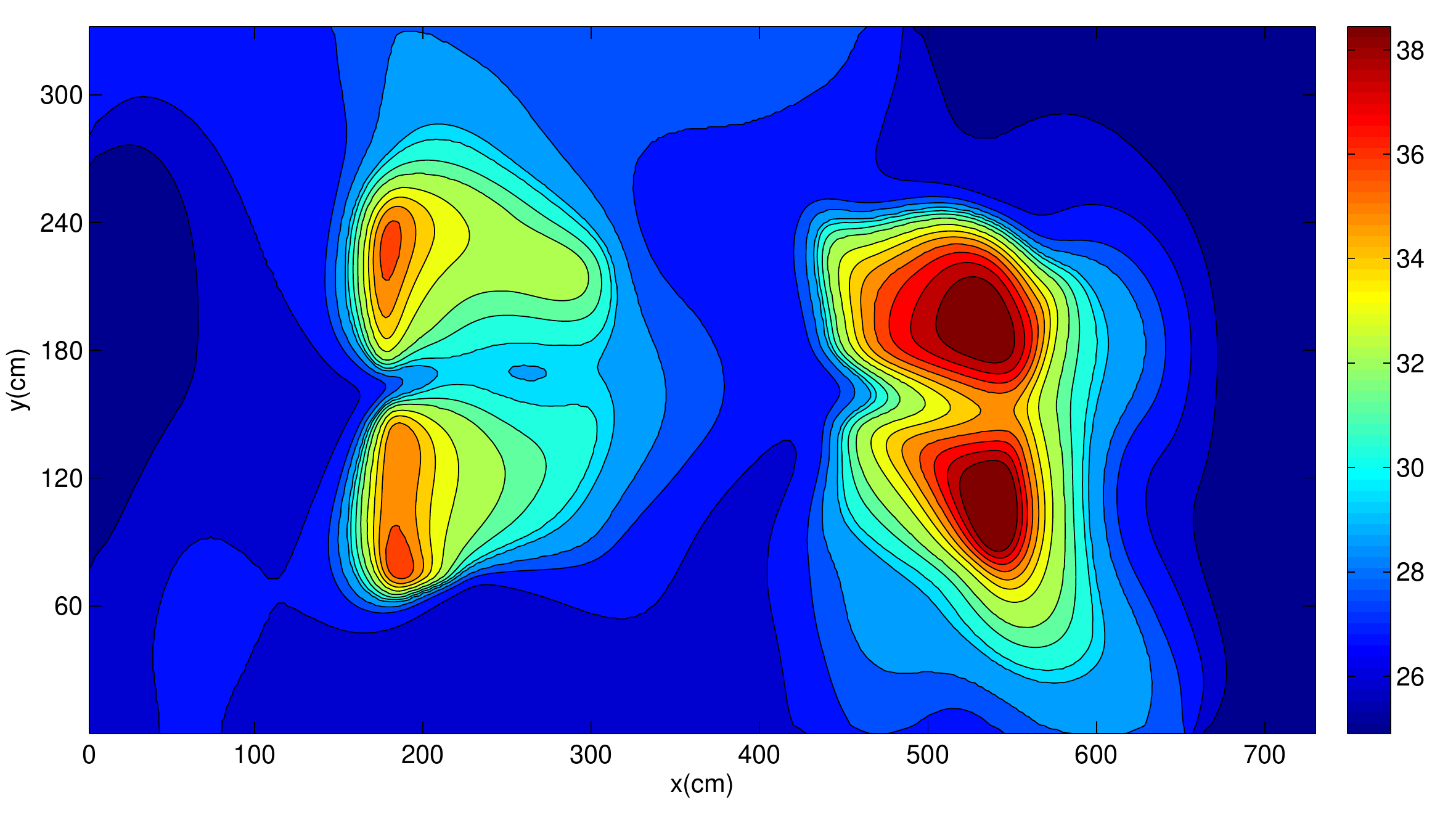}
 \caption{the thermal map obtained from CFD simulation}
    \label{Fig:cfdThermalMap}
    \end{subfigure}%

    \begin{subfigure}[b]{0.75\textwidth}
        \centering
        \includegraphics[width=\textwidth]{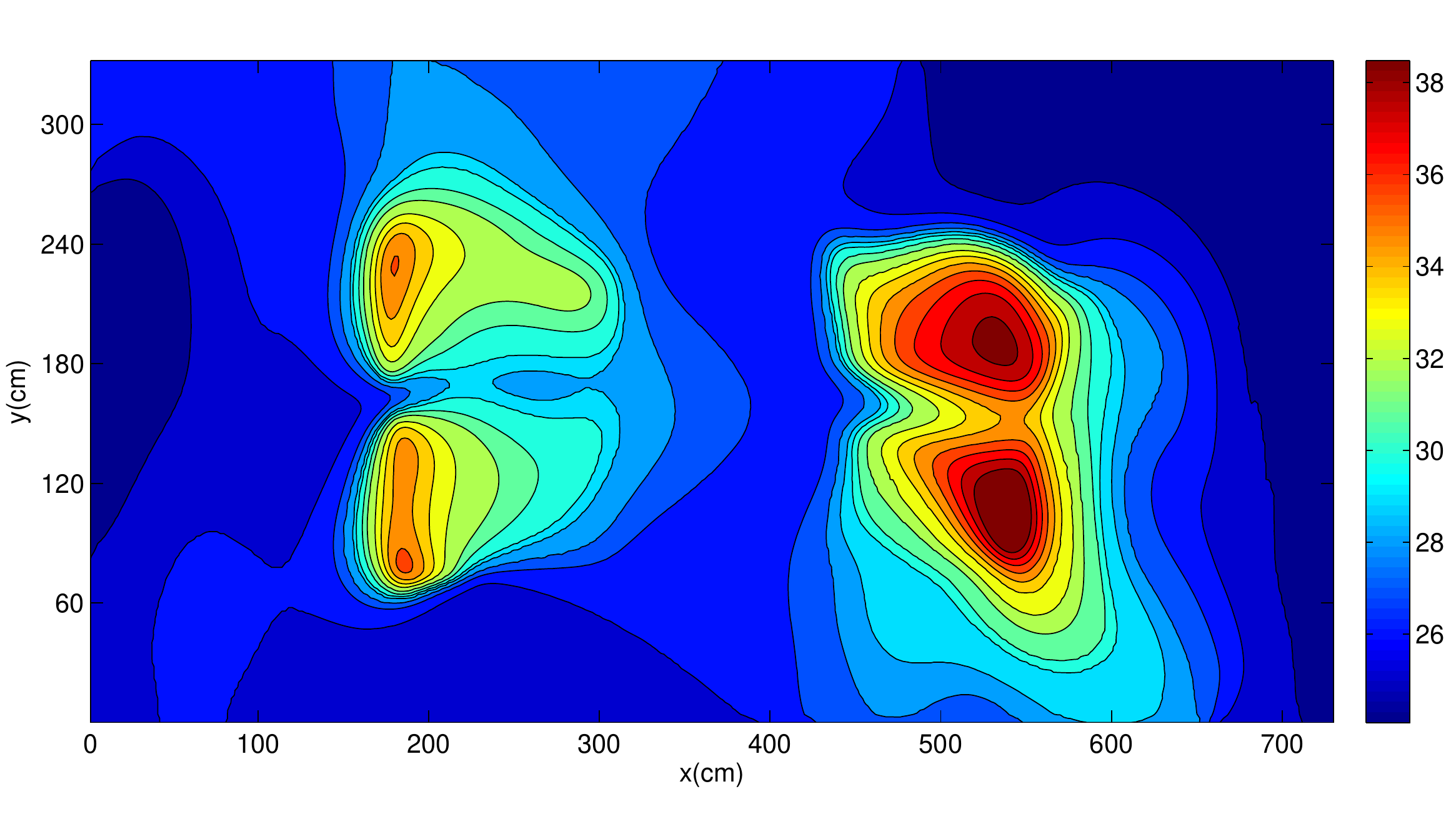}
       \caption{the calibrated thermal map with $\alpha=0.01, \sigma_m=1000$ and $\sigma_d=1$}
    \label{Fig:calThermalMap}
    \end{subfigure}
 \caption{The thermal map (temperature field in $^\circ C$) of the room at 1.2m height.}
    \label{Fig:ThermalMap}
\end{figure}

The convergence criteria were set as $10^{-6}$ for energy and $10^{-3}$ for all the other parameters. As shown in Fig. \ref{Fig:converge}, the simulation converged with 442 iterations. We are interested in the thermal map of the room at 1.2m height. The thermal map obtained from the CFD simulation is shown in Fig. \ref{Fig:cfdThermalMap}.

\subsection{The calibration results}

We use the sensor observations obtained from sensors s1-s4 to calibrate the thermal map in Fig. \ref{Fig:cfdThermalMap}. The four sensors s1-s4 were just above the centers of the four heaters, respectively. 

We set $\sigma_m=1000$ and $\sigma_d=1$.
The thermal map in Fig. \ref{Fig:cfdThermalMap} consists of 60955$(=167\times 365)$ temperature values of all the mesh points at the plane $z=1.2$m. The balance factor $\lambda$ is related to the number of mesh points and the number of sensor observations. We set
\begin{equation}\label{eq:lambda}
  \lambda = \alpha\frac{m}{N}=\frac{4\alpha}{60955}
\end{equation}
where $0<\alpha\leq1$. The first part in \eqref{eq:Japprox} is the summation of $m\times N$ terms while the second part is the summation of $N\times N$ terms. If we set $\alpha = 1$, it means that we have the same trust on the sensor observations and the simulation results on all mesh points. Obviously, sensor observations are more accurate. We set $\alpha=0.01$ and the calibrated thermal map is shown in Fig. \ref{Fig:calThermalMap}. Here, $\alpha=0.01$ implies that the influence of each sensor observation in \eqref{eq:Japprox} is 100 times important than the simulated data on each mesh point.

\begin{figure*}[tb]
    \centering
    \begin{subfigure}[b]{0.45\textwidth}
        \centering
        \includegraphics[width=\textwidth]{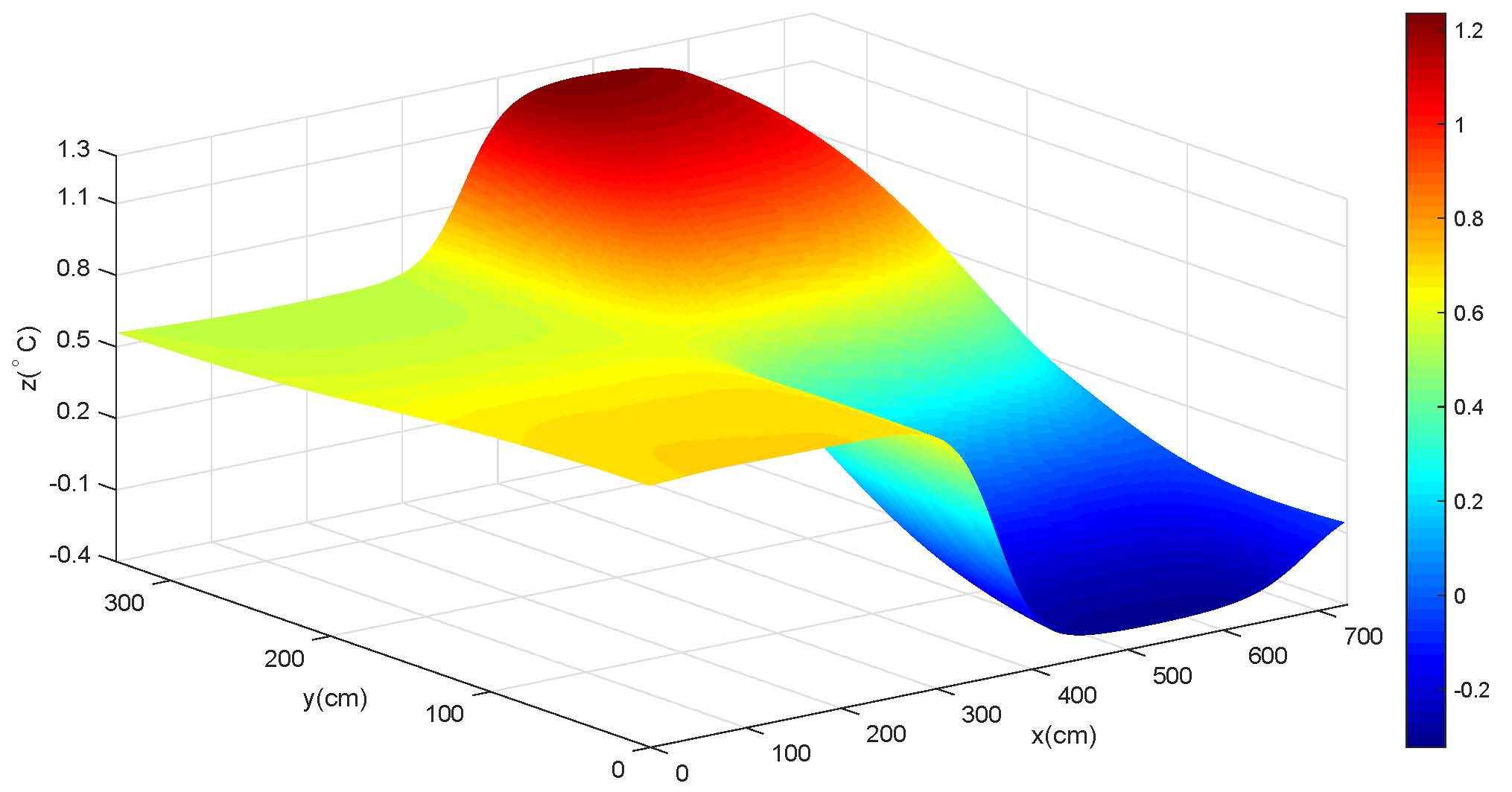}
        \subcaption{$\alpha = 0.01$}
        \label{Fig:ErrorMap1}
    \end{subfigure}
    \begin{subfigure}[b]{0.45\textwidth}
        \centering
        \includegraphics[width=\textwidth]{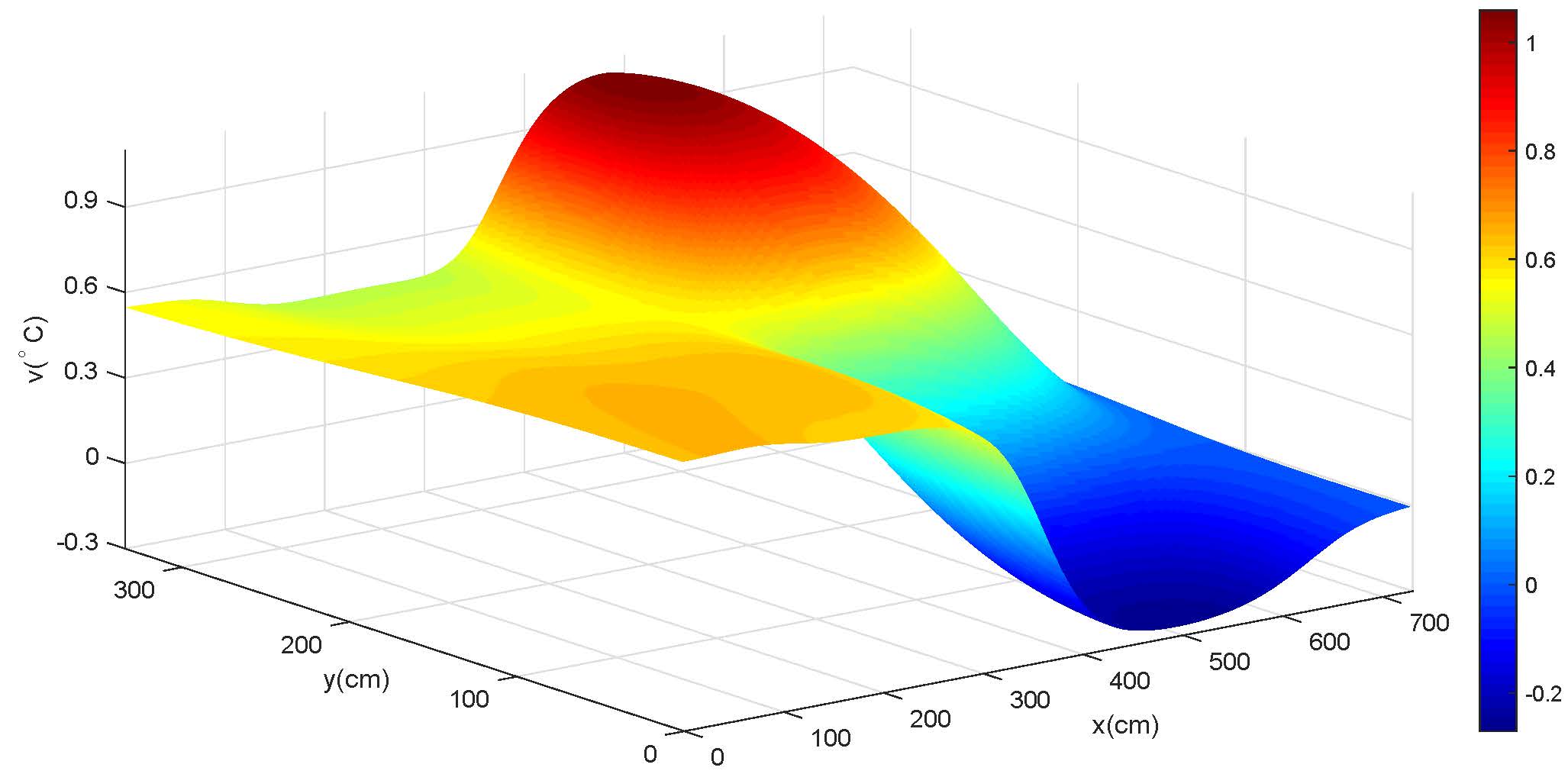}
        \subcaption{$\alpha = 0.1$}
    \end{subfigure}
        \begin{subfigure}[b]{0.45\textwidth}
        \centering
        \includegraphics[width=\textwidth]{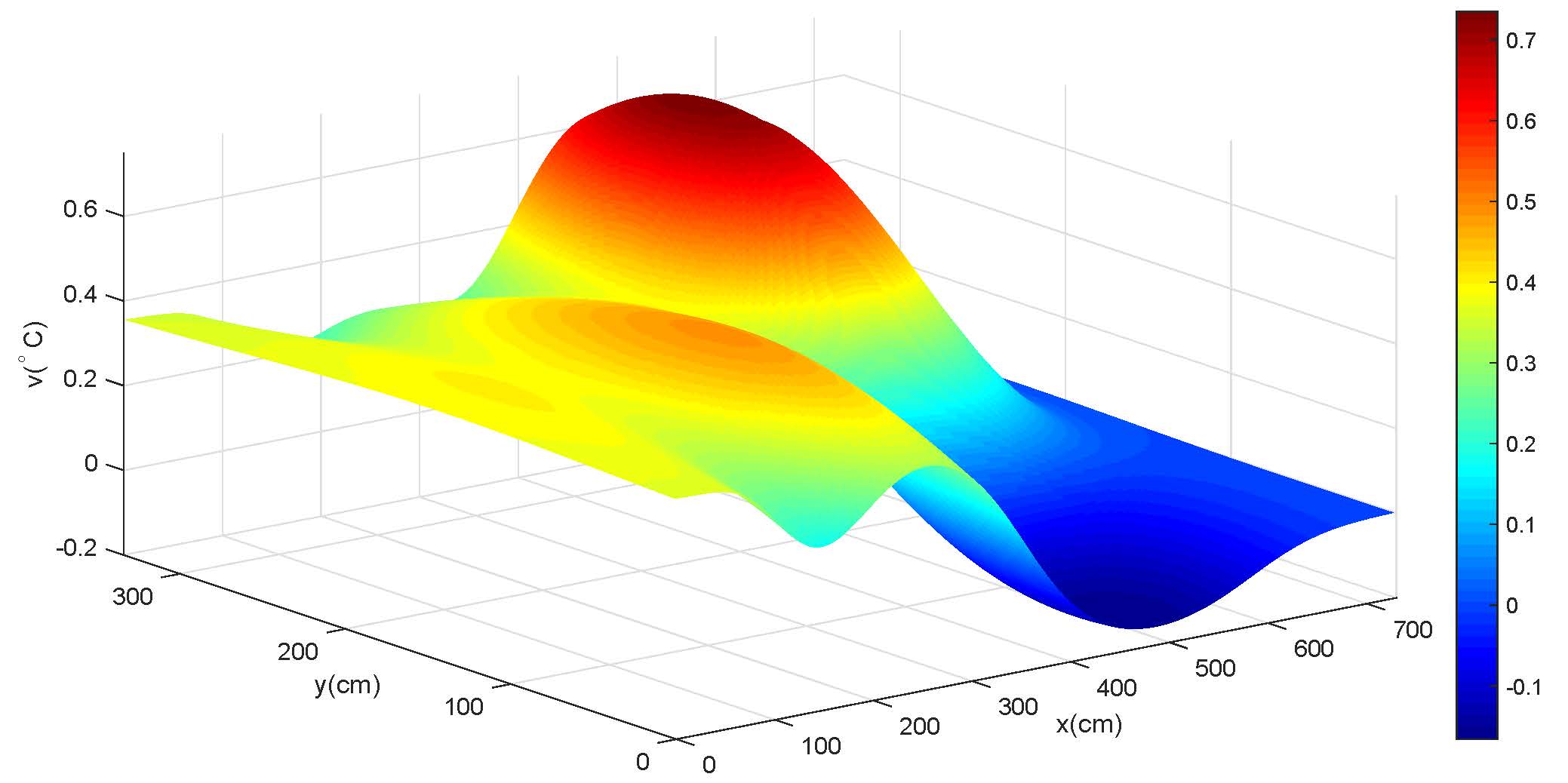}
        \subcaption{$\alpha = 0.5$}
    \end{subfigure}
        \begin{subfigure}[b]{0.45\textwidth}
        \centering
        \includegraphics[width=\textwidth]{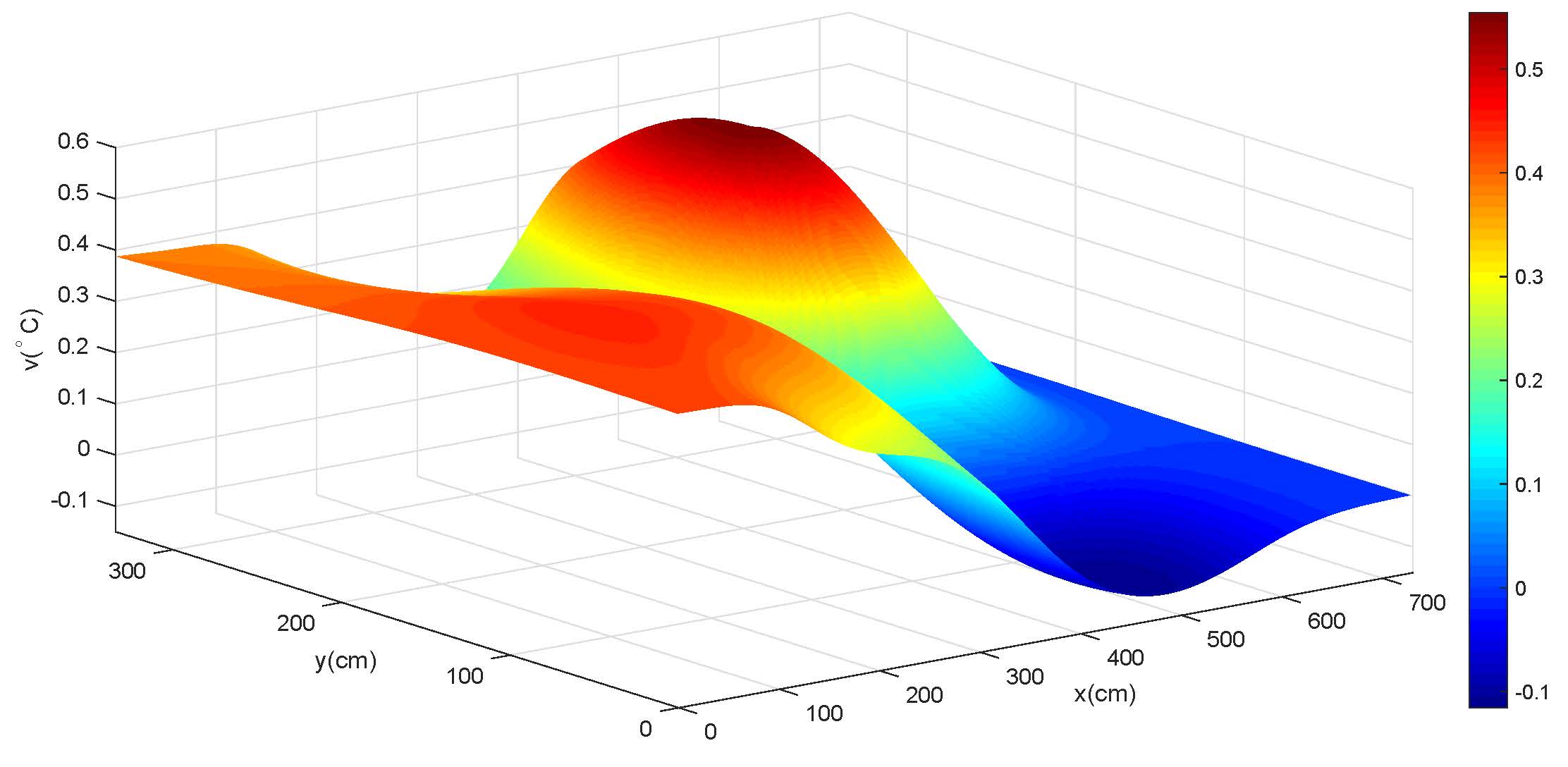}
        \subcaption{$\alpha = 1$}
    \end{subfigure}
 \caption{The estimated error ($\hat{\mathbf{v}}$) of the thermal map in Fig. \ref{Fig:cfdThermalMap} w.r.t. different balance factor. Both $\sigma_m$ and $\sigma_d$ are fixed and set as $\sigma_m=1000$ and $\sigma_d=1$.}
    \label{Fig:EstimatedError}
\end{figure*}

Comparing Fig. \ref{Fig:calThermalMap} with Fig. \ref{Fig:cfdThermalMap}, we can easily find that the calibrated thermal map preserves the profile of the CFD results. However, Fig.\ref{Fig:calThermalMap} cannot show how the simulated thermal map is calibrated. We show the estimated error (i.e., $\hat{\mathbf{v}}$) in Fig. \ref{Fig:ErrorMap1}, which is the difference between Fig. \ref{Fig:cfdThermalMap} and Fig. \ref{Fig:calThermalMap}. Considering \eqref{eq:fcalibrated}, we can obtain the simulated thermal map by subtracting the estimated error from the simulated thermal map.

\subsection{The influence of balance factor}

\begin{figure}[htb]
    \centering
        \includegraphics[width=0.7\textwidth]{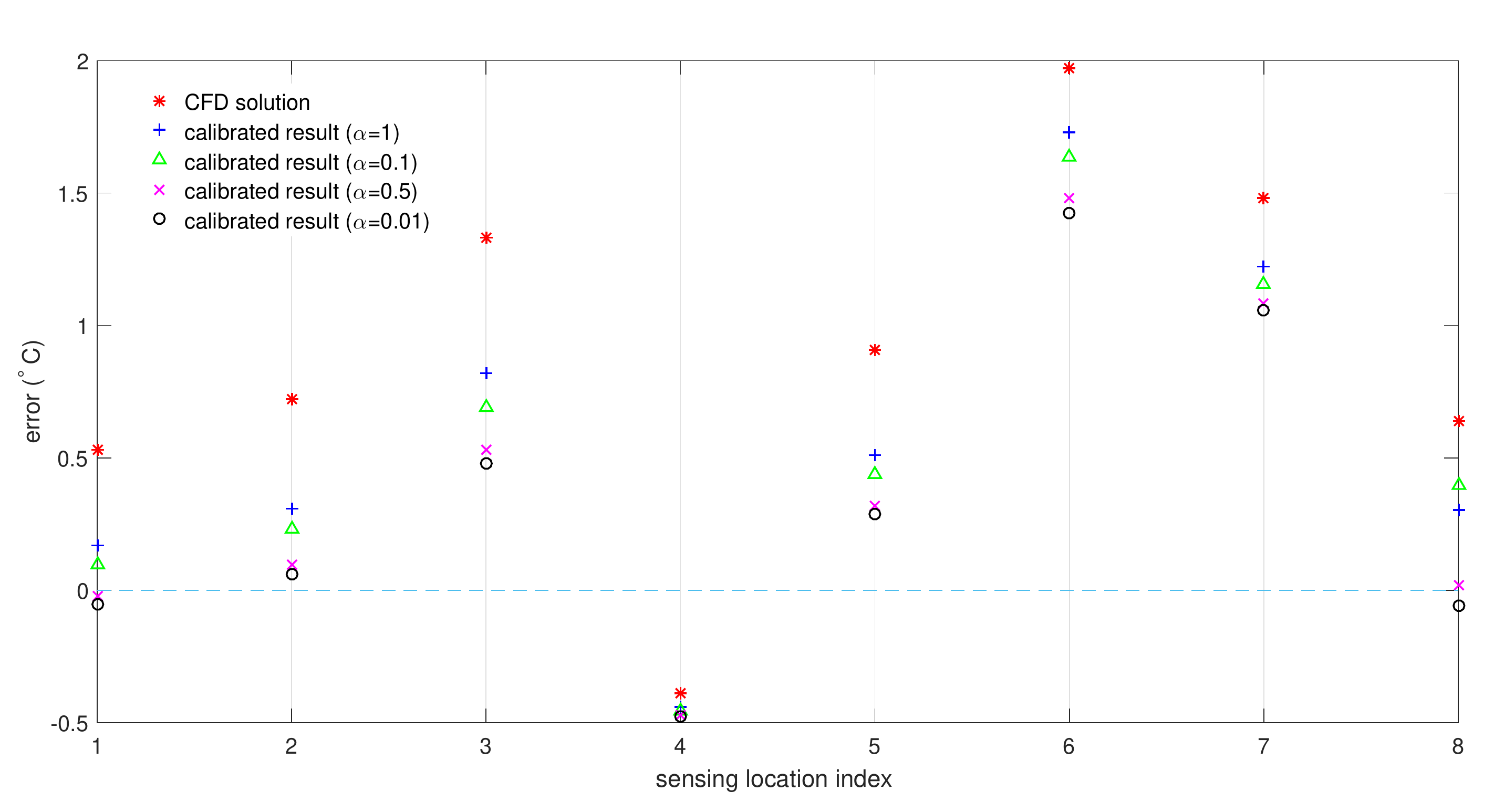}
 \caption{The errors of CFD solution and the calibrated thermal maps at the sensing locations s1-s8 with $\sigma_m=1000$ and $\sigma_d=1$}
    \label{Fig:SensorError}
\end{figure}

To check the effect of the balance factor, we set $\alpha$ as 0.01, 0.1, 0.5, and 1 respectively. The corresponding estimated errors of the thermal map in Fig. \ref{Fig:cfdThermalMap} is given in Fig \ref{Fig:EstimatedError}. The observations of sensors s1-s4 are used to estimate the error.
To show the effectiveness of the proposed method, we use the observations of sensor s5-s8 as the ground truth to test the estimation performance. The error of the CFD simulation results and the calibrated thermal map at the 8 sensing locations are shown in Fig. \ref{Fig:SensorError}.

Fig. \ref{Fig:EstimatedError} shows that with different balance factors, the estimated errors (i.e., $\hat{\mathbf{v}}$) almost have the same profiles. However, the magnitude of the estimated errors decrease when the balance factor increases, which implies that the influence of sensor observations decreases and we can only provide a very small adjustment for the original simulated thermal map. In other words, the balance factor can control the magnitude of the estimated errors, i.e., the magnitude of the adjustment for the simulated thermal map. As shown in Fig. \ref{Fig:EstimatedError} and Fig. \ref{Fig:SensorError}, with a smaller balance factor, the magnitude of the estimated error is larger and we can obtain a better performance.

\begin{table*}[tbh]
\caption{The RMSE of the thermal maps data at the sensing locations w.r.t. different balance factor}
\begin{center}
\begin{tabular}{l|ll|ll}
\hline
Data at the sensing locations & RMSE(s1-s8) & improvement(s1-s8) & RMSE(s5-s8) & improvement(s5-s8)\\
\hline
CFD simulation result           & 1.1197     & /         &1.3518 & /  \\
calibrated result ($\alpha=1$) & 0.8536      & 23.77\%   &1.0999 & 18.55\%\\
calibrated result ($\alpha=0.5$) & 0.7996    & 28.59\%   &1.0451 & 22.59\%\\
calibrated result ($\alpha=0.1$) & 0.7045    & 37.08\%   &0.9301 & 31.06\%\\
calibrated result ($\alpha=0.01$) & 0.6794   & 39.32\%   &0.8985 & 33.38\%\\
\hline\hline
\end{tabular}
\begin{tablenotes}
Note: Both $\sigma_m$ and $\sigma_d$ are fixed. We set $\sigma_m=1000$ and $\sigma_d=1$. Here,
$\emph{improvement} = (\emph{RMSE of CFD simulation result} - \emph{RMSE of calibrated result})/\emph{RMSE of CFD simulation result}$.
\end{tablenotes}
\label{Table:RMSE}
\end{center}
\end{table*}

The root mean square error (RMSE) of the thermal map data at sensing locations are given in Table \ref{Table:RMSE}.
Both Fig. \ref{Fig:SensorError} and Table \ref{Table:RMSE} show that we can enhance the thermal map obtained from CFD simulation with the four sensor observations (s1-s4).
With the calibration, the RMSE of the data at sensing locations s5-s8 can be improved by $33.38\%$.
However,  we can see from Fig. \ref{Fig:SensorError} that the errors of the regions around s6 and s7 are still large.

The CFD simulation results around s6 and s7 have large errors, i.e., around $2^\circ C$ and $1.5^\circ C$, respectively. The proposed method reduces the errors to around $1.5^\circ C$ and $1^\circ C$, which are still large. To address this issue, we should consider the capacity of the proposed method. The largest magnitude to which the proposed method can adjust the CFD results almost equals the maximum error of the CFD results at all sensing locations where the sensor observations were used to calibrate the CFD results.
It can be seen in Fig. \ref{Fig:SensorError} that the maximum error of the CFD results at sensing locations s1-s4 is around $1.3^\circ C$. Therefore, the maximum value of the estimated error of the simulated thermal map is less than $1.3^\circ C$. The balance factor can control the magnitude of the maximum value of the estimated errors. In other words, the balance factor can control the magnitude of the adjustment for the simulated thermal map, but the upper bound of this magnitude is determined by the sensing locations.

\begin{figure*}[tb]
    \centering
    \begin{subfigure}[b]{0.45\textwidth}
        \centering
        \includegraphics[width=\textwidth]{eMap1per.jpg}
        \subcaption{$\sigma_m=1000$}
    \end{subfigure}
    \begin{subfigure}[b]{0.45\textwidth}
        \centering
        \includegraphics[width=\textwidth]{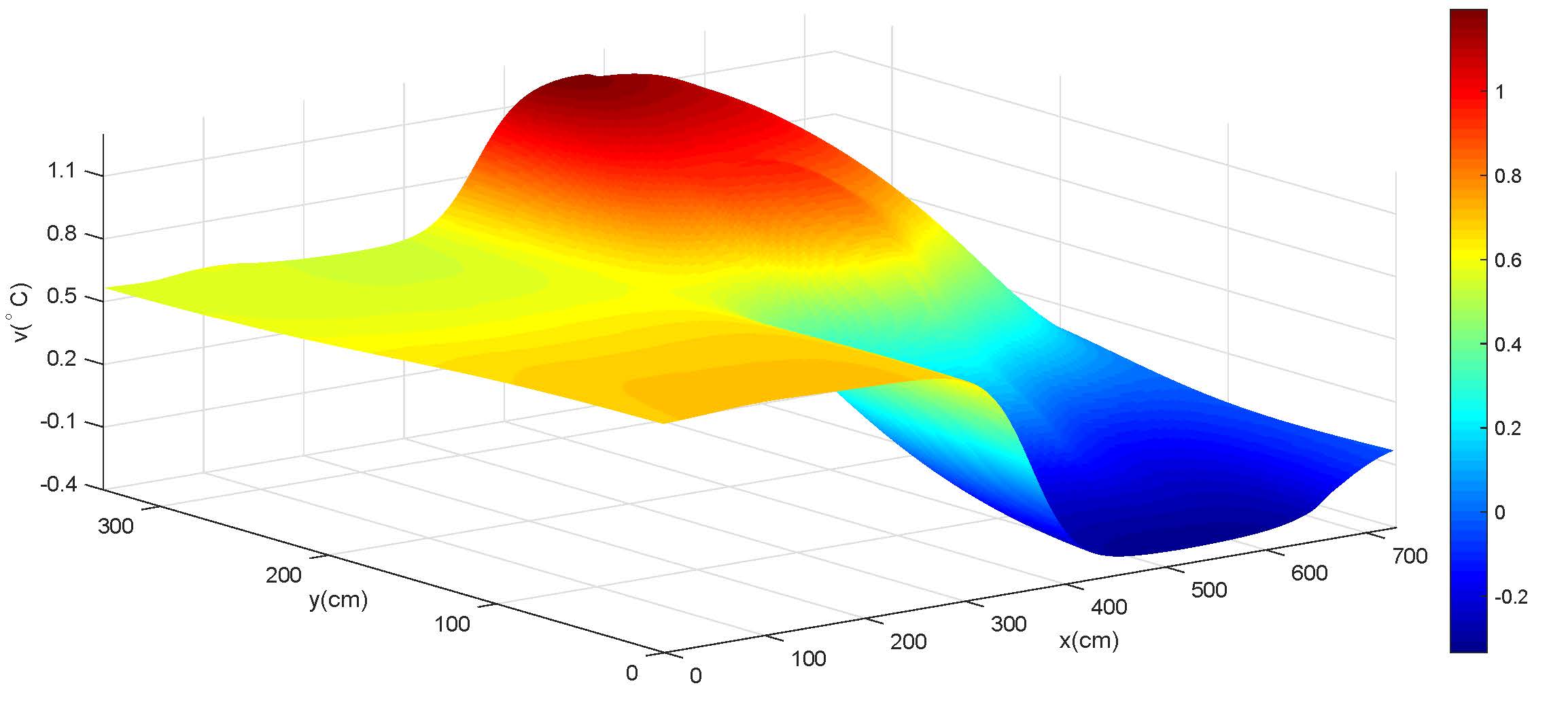}
        \subcaption{$\sigma_m=100$}
    \end{subfigure}
        \begin{subfigure}[b]{0.45\textwidth}
        \centering
        \includegraphics[width=\textwidth]{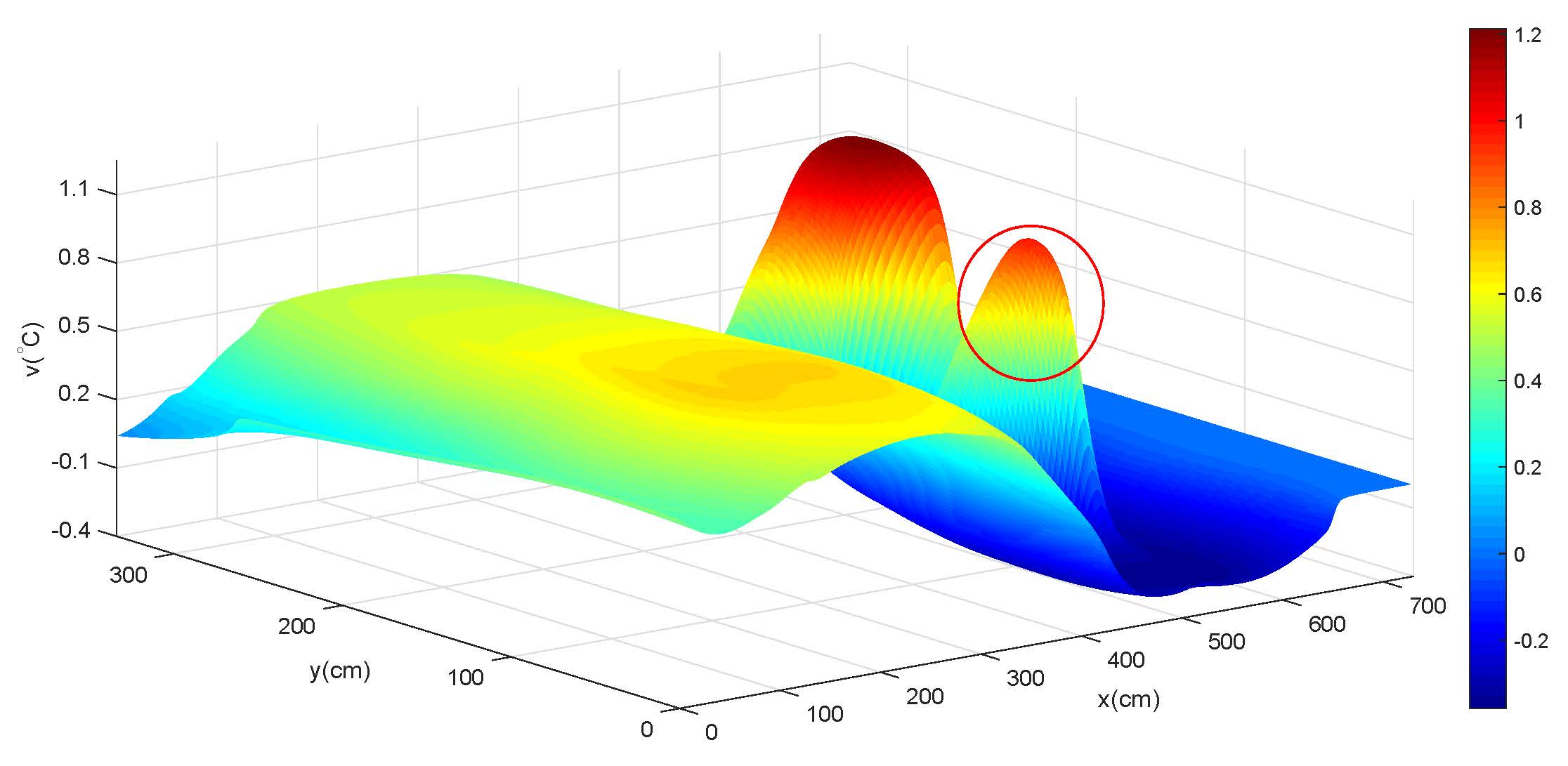}
        \subcaption{$\sigma_m=10$}
        \label{Fig:ErrorMVc}
    \end{subfigure}
        \begin{subfigure}[b]{0.45\textwidth}
        \centering
        \includegraphics[width=\textwidth]{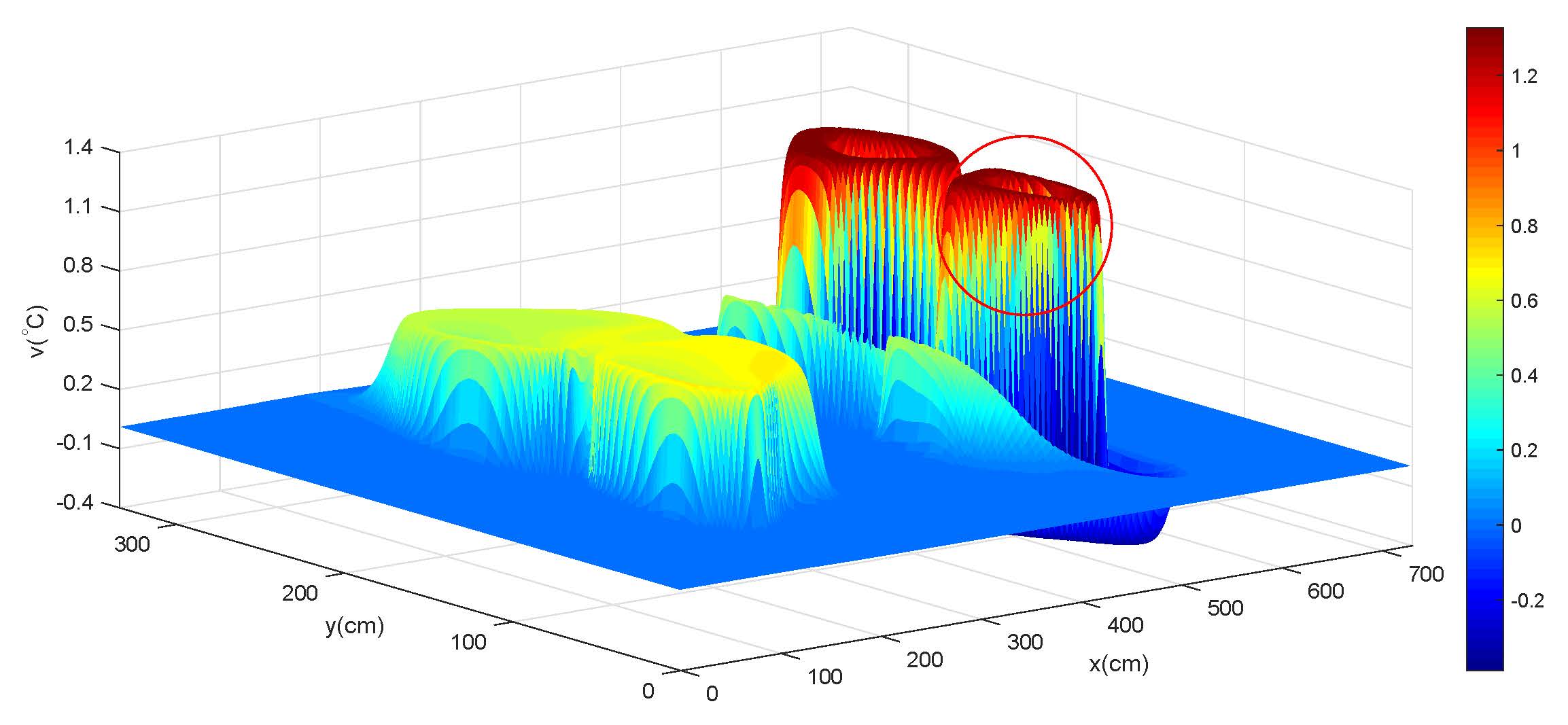}
        \subcaption{$\sigma_m=1$}
        \label{Fig:ErrorMVd}
    \end{subfigure}
 \caption{The estimated error ($\hat{\mathbf{v}}$) of the thermal map in Fig. \ref{Fig:cfdThermalMap} w.r.t. different magnitude variance $\sigma_m$. Both $\alpha$ and $\sigma_d$ are fixed and set as $\alpha=0.01$ and $\sigma_d=1$.}
    \label{Fig:ErrorMV}
\end{figure*}

\subsection{The influence of magnitude variance}

\begin{figure}[htb]
    \centering
        \includegraphics[width=0.7\textwidth]{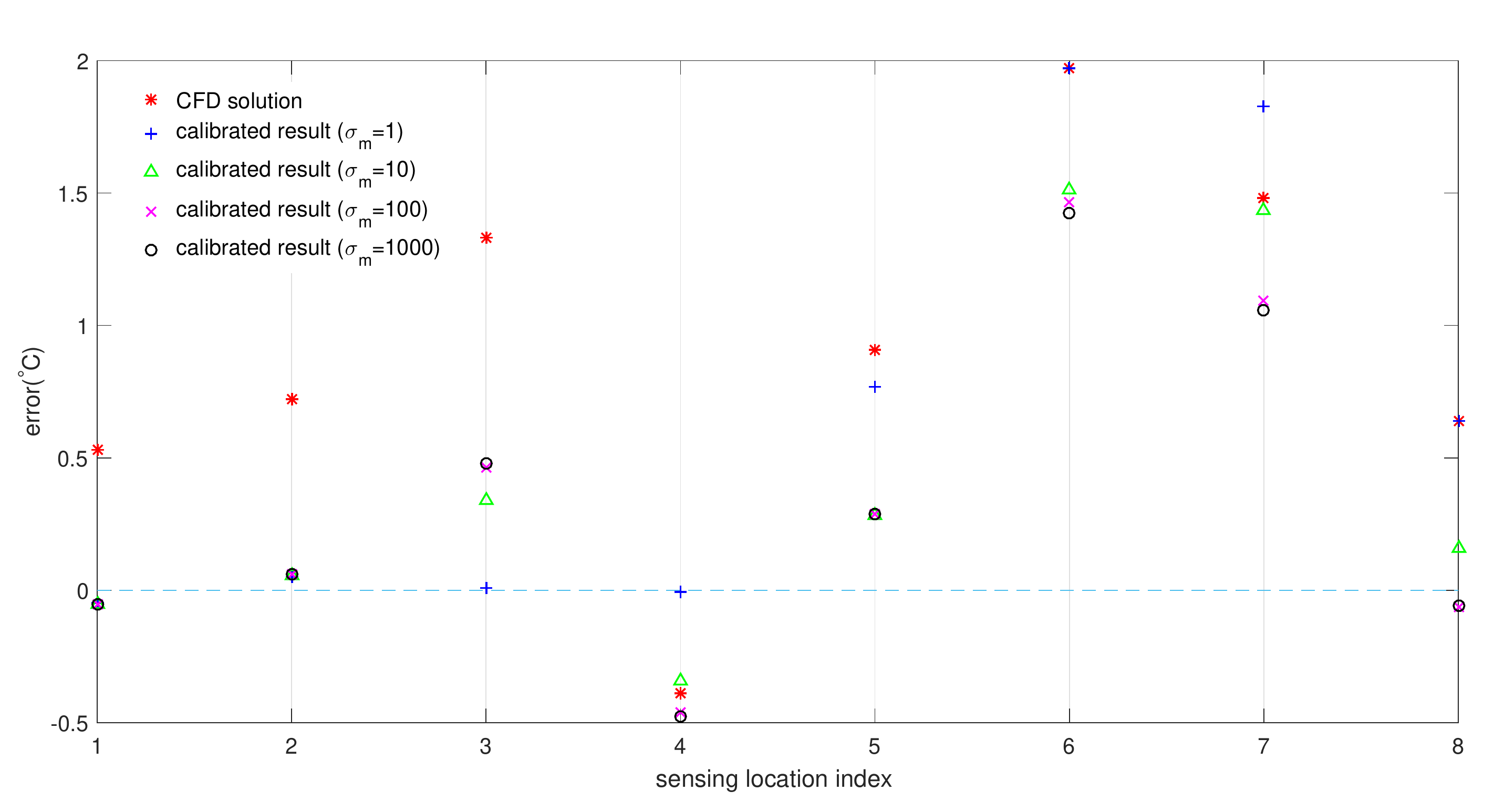}
 \caption{The errors of CFD simulation result and the calibrated thermal maps at the sensing locations s1-s8 with $\alpha=0.01$ and $\sigma_d=1$}
    \label{Fig:SensorErrorM}
\end{figure}

To check the influence of magnitude variance, we set $\sigma_m$ as  1000, 100, 10, and 1, respectively. With the four sensor observations (s1-s4), we estimated the error of the thermal map in Fig. \ref{Fig:cfdThermalMap}. The estimated errors are given in Fig.\ref{Fig:ErrorMV}. The error of the estimated thermal map at the 8 sensing locations are shown in Fig. \ref{Fig:SensorErrorM}, and the RMSE of the calibrated thermal map at the 8 sensing locations are given in Table \ref{Table:RMSEM}.

Fig. \ref{Fig:ErrorMV} shows that with different magnitude variances, the maximum magnitude of the estimated errors (i.e., $\hat{\mathbf{v}}$) are almost the same, which implies that different with the balance factor, the magnitude variance has no significant influence on the the maximum magnitude of the estimated errors.

It can be seen from Fig. \ref{Fig:ErrorMVc}-\ref{Fig:ErrorMVd} that the region in the two red circle is nearer to sensor s4 than s3 but the estimated error is similar with the estimated error around s3. We can see form Fig. \ref{Fig:cfdThermalMap} that the temperature values of the thermal map in the region marked by the two circles (in Fig. \ref{Fig:ErrorMVc}-\ref{Fig:ErrorMVd}) are similar to those in the region around sensor s3. The proposed thermal map calibration work is a process of fusing sensor information and the information of CFD results.
A smaller magnitude variance can enhance the influence of the CFD results in this information fusion work.

In addition, Fig. \ref{Fig:ErrorMVd} shows that if $\sigma_m=1$ in the regions far away the sensing locations s1-s4, the estimated errors are almost zero. In other words, the proposed method has no influence for these regions. As we mentioned before, a small magnitude variance leads to a small value of the first Gaussian function in \eqref{eq:wight}, which can dilute the influence of the second Gaussian function. The smaller the magnitude variance $\sigma_m$, the smaller the weight $w(\mathbf{x}_k, \mathbf{y})$, especially when $\mathbf{y}$ is far away the sensing location $\mathbf{x}_k$. Therefore, if the magnitude variance is too small, the proposed method could not globally calibrated the thermal map.

\begin{table*}[tbh]
\caption{The RMSE of the thermal maps data at the sensing locations w.r.t. different magnitude variance $\sigma_m$}
\begin{center}
\begin{tabular}{l|ll|ll}
\hline
Data at the sensing locations & RMSE(s1-s8) & improvement(s1-s8) & RMSE(s5-s8) & improvement(s5-s8)\\
\hline
CFD simulation result           & 1.1197     & /         &1.3518 & /  \\
calibrated result ($\sigma_m=1$) & 1.0142      & 9.42\%   &1.4339 & -6.05\%\\
calibrated result ($\sigma_m=10$) & 0.7652    & 31.66\%   &1.0543 & 21.91\%\\
calibrated result ($\sigma_m=100$) & 0.6957    & 37.87\%   &0.9267 & 31.31\%\\
calibrated result ($\sigma_m=1000$) & 0.6794   & 39.32\%   &0.8985 & 33.38\%\\
\hline\hline
\end{tabular}
\begin{tablenotes}
Note: Both $\alpha$ and $\sigma_d$ are fixed. We set $\alpha=0.01$ and $\sigma_d=1$. Here,
$\emph{improvement} = (\emph{RMSE of CFD simulation result} - \emph{RMSE of calibrated result})/\emph{RMSE of CFD simulation result}$.
\end{tablenotes}
\label{Table:RMSEM}
\end{center}
\end{table*}

Fig. \ref{Fig:SensorErrorM} shows that with the decrease of the magnitude variance, the errors around sensing locations s1-s4 also decrease but those around sensing locations s5-s8 do not. From Table \ref{Table:RMSEM} we can find that the RMSE at sensing locations s5-s8 increases with respect to the decrease of the magnitude variance. If we reduce the magnitude variance to 1, the calibration work makes the accuracy of the simulated thermal map worse. Hence, we conclude that reducing the magnitude variance can improve the local calibration but has no significant help for global calibration. If enough sensor observations are available we can set a small magnitude variance unless a large one is proper.

\subsection{The influence of distance variance}

\begin{figure*}[tb]
    \centering
    \begin{subfigure}[b]{0.45\textwidth}
        \centering
        \includegraphics[width=\textwidth]{eMap1per.jpg}
        \subcaption{$\sigma_d=1$}
    \end{subfigure}
    \begin{subfigure}[b]{0.45\textwidth}
        \centering
        \includegraphics[width=\textwidth]{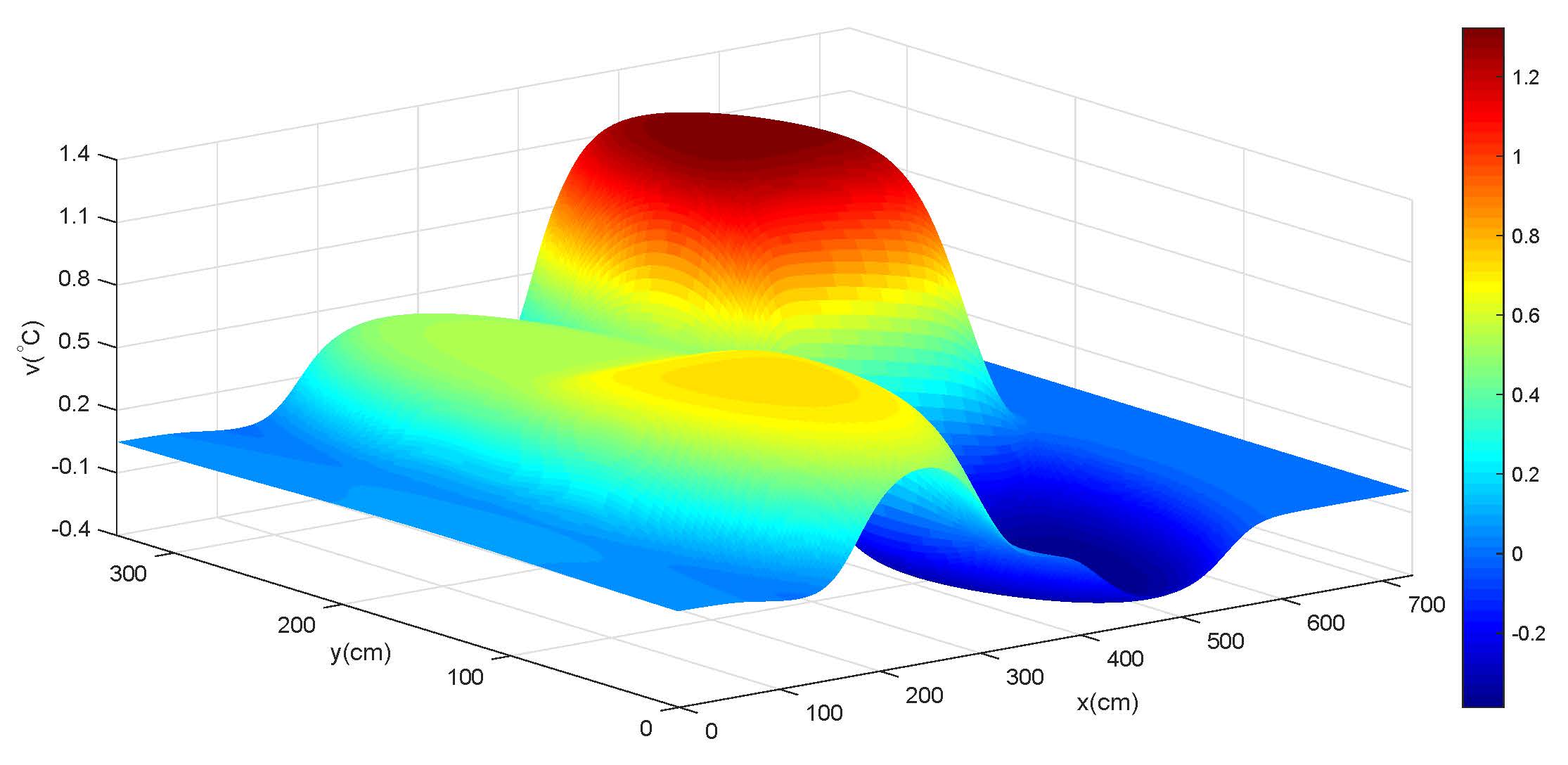}
        \subcaption{$\sigma_d=0.5$}
    \end{subfigure}
        \begin{subfigure}[b]{0.45\textwidth}
        \centering
        \includegraphics[width=\textwidth]{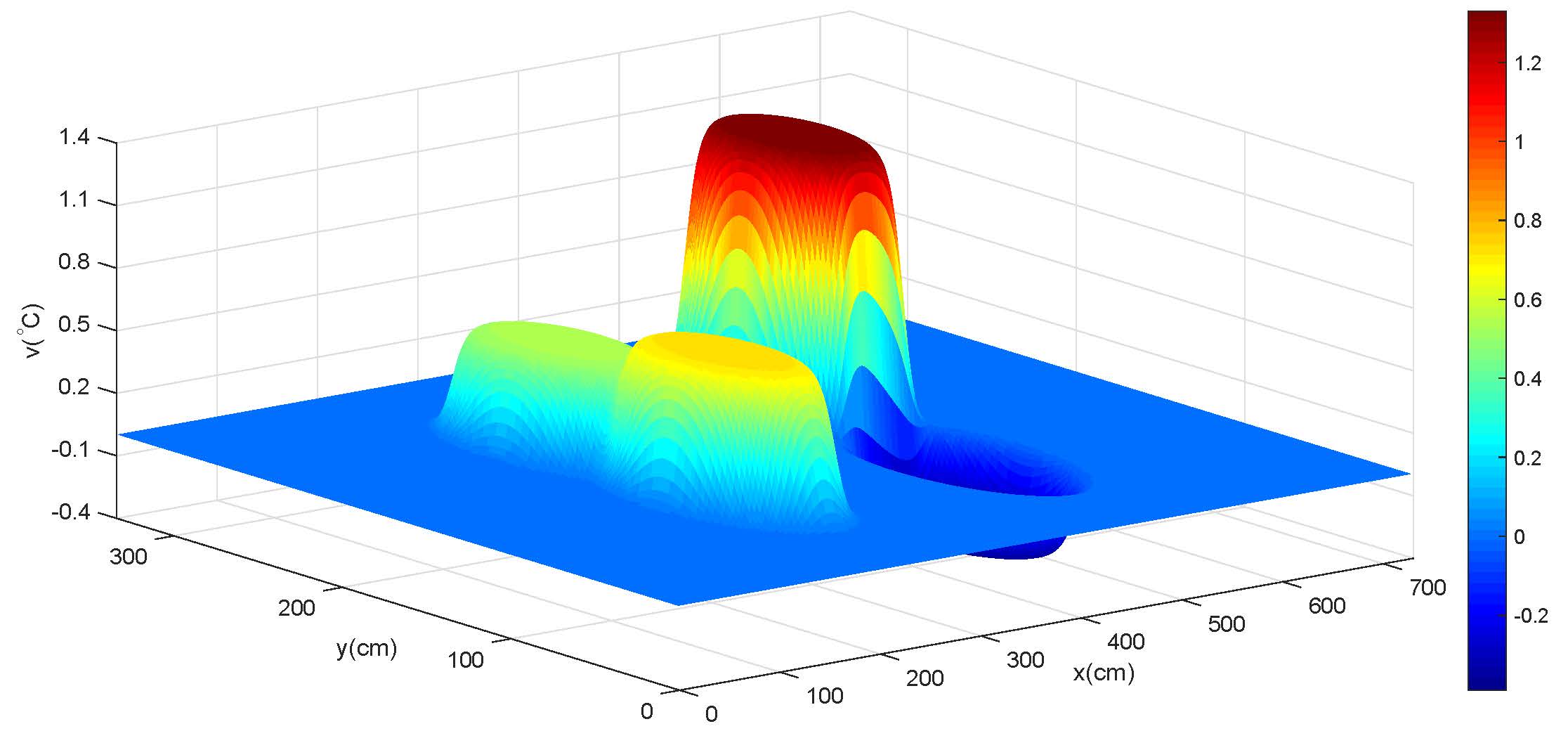}
        \subcaption{$\sigma_d=0.2$}
    \end{subfigure}
        \begin{subfigure}[b]{0.45\textwidth}
        \centering
        \includegraphics[width=\textwidth]{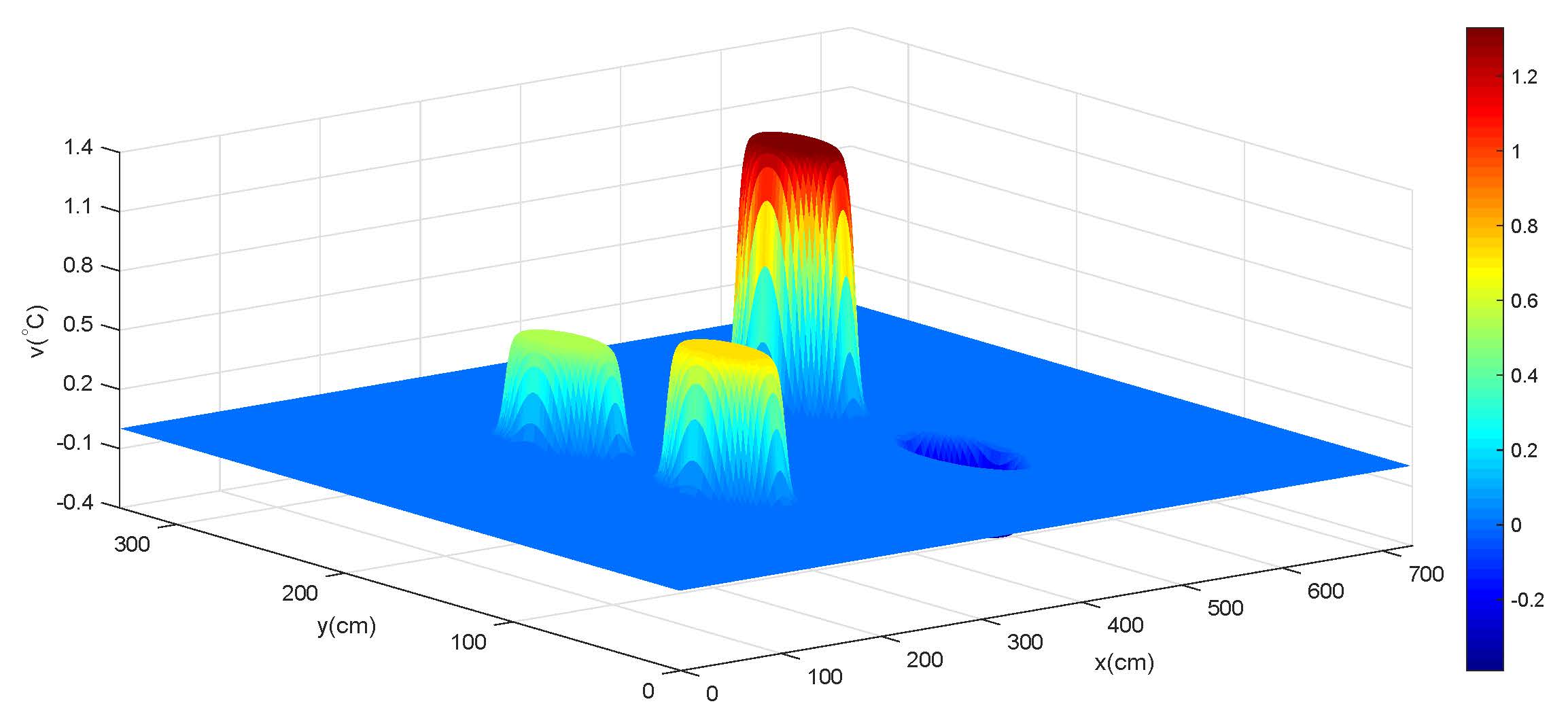}
        \subcaption{$\sigma_d=0.1$}
    \end{subfigure}
 \caption{The estimated error ($\hat{\mathbf{v}}$) of the thermal map in Fig. \ref{Fig:cfdThermalMap} w.r.t. different distance variance $\sigma_d$. Both $\alpha$ and $\sigma_m$ are fixed and set as $\alpha=0.01$ and $\sigma_m=1000$.}
    \label{Fig:ErrorDV}
\end{figure*}

\begin{figure}[htb]
    \centering
        \includegraphics[width=0.7\textwidth]{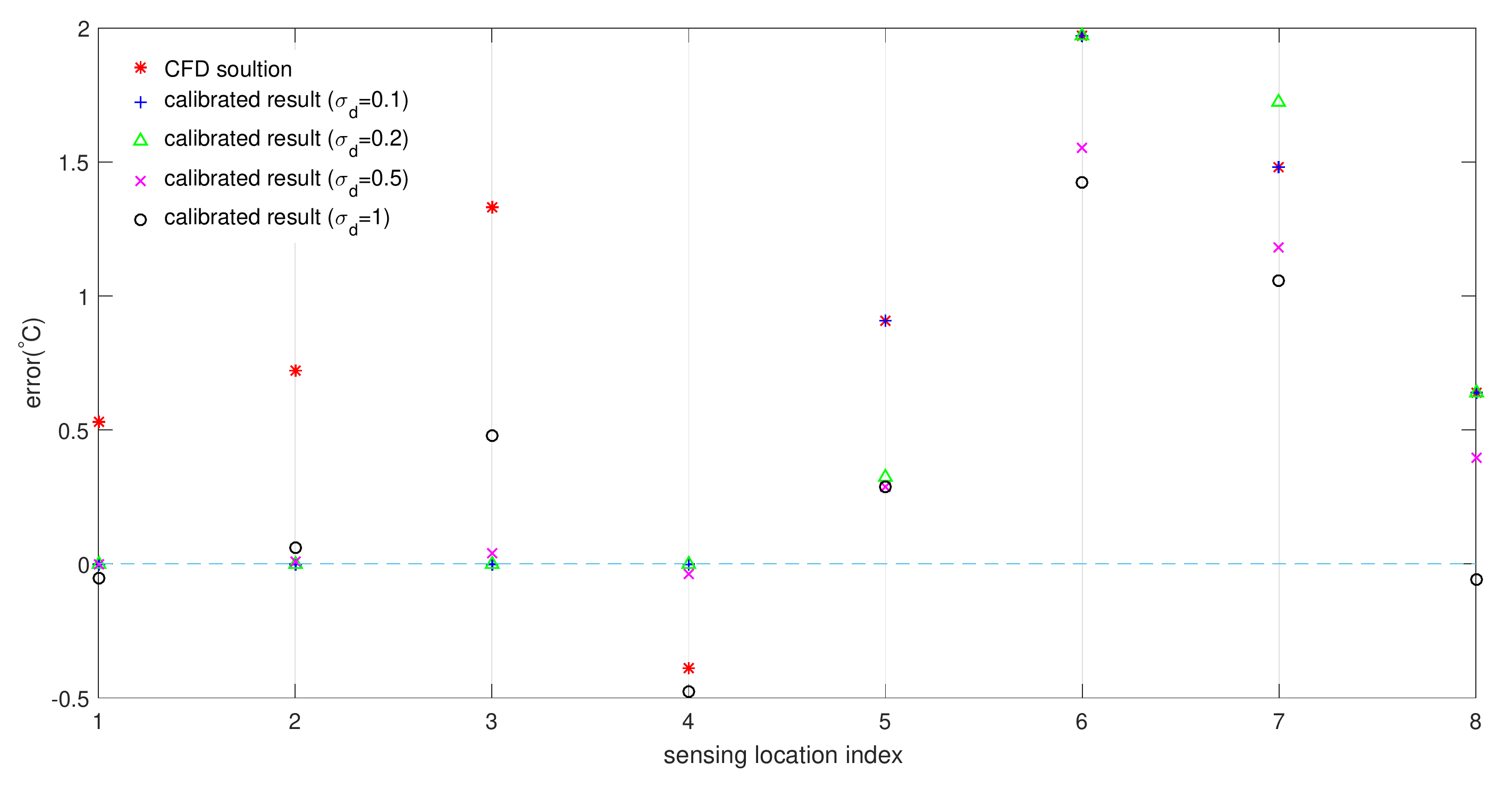}
 \caption{The errors of CFD simulation result and the calibrated thermal maps at the sensing locations s1-s8 with $\alpha=0.01$ and $\sigma_m=1000$}
    \label{Fig:SensorErrorD}
\end{figure}

To check the influence of distance variance, we set $\sigma_d$ as  1, 0.5, 0.2, and 0.1, respectively. With the four sensor observations (s1-s4), we estimated the error of the thermal map in Fig. \ref{Fig:cfdThermalMap}. The estimated errors are given in Fig.\ref{Fig:ErrorDV}. The error of the estimated thermal map at the 8 sensing locations are shown in Fig. \ref{Fig:SensorErrorD}, and the RMSE of the calibrated thermal map at the 8 sensing locations are given in Table \ref{Table:RMSED}.

\begin{table*}[tbh]
\caption{The RMSE of the thermal maps data at the sensing locations w.r.t. different magnitude variance $\sigma_m$}
\begin{center}
\begin{tabular}{l|ll|ll}
\hline
Data at the sensing locations & RMSE(s1-s8) & improvement(s1-s8) & RMSE(s5-s8) & improvement(s5-s8)\\
\hline
CFD simulation result           & 1.1197     & /         &1.3518 & /  \\
calibrated result ($\alpha=1$)  & 0.9558      & 14.64\%   &1.3518 & 0\%\\
calibrated result ($\alpha=0.5$) & 0.9594    & 14.32\%   &1.3568 & -0.37\%\\
calibrated result ($\alpha=0.1$) & 0.7117    & 36.44\%   &1.0062 & 25.45\%\\
calibrated result ($\alpha=0.01$) & 0.6794   & 39.32\%   &0.8985 & 33.38\%\\
\hline\hline
\end{tabular}
\begin{tablenotes}
Note: Both $\alpha$ and $\sigma_d$ are fixed. We set $\alpha=0.01$ and $\sigma_d=1$. Here,
$\emph{improvement} = (\emph{RMSE of CFD simulation result} - \emph{RMSE of calibrated result})/\emph{RMSE of CFD simulation result}$.
\end{tablenotes}
\label{Table:RMSED}
\end{center}
\end{table*}

From Fig. \ref{Fig:ErrorDV} we can find that like the magnitude variance, the distance variance has no significant influence on the maximum value of the estimated errors. It is clear shown in Fig. \ref{Fig:ErrorDV} that with a smaller distance variance, the scope of non-zero region of the estimated errors dramatically decrease. Fig. \ref{Fig:SensorErrorD} shows that when the distance variance $\sigma_d\leq0.5$, the proposed method can provide almost perfect calibration at the four sensing locations (s1-s4). Table \ref{Table:RMSED} shows that with a smaller distance variance, the errors of the calibrated thermal map at the sensing locations s5-s8 increase.  These agree with our previous analysis that the distance variance controls the scope of the region on which the sensor observation has significant influence. A small distance variance $\sigma_d$ can lead to very good local calibration while a large $\sigma_d$ can provide global influence for an isolated sensor observation.

\section{concluding remarks}
CFD simulation results calibration from sparse sensor observations is a new and interesting problem. We formulated this problem as an optimization problem. The proposed method can provide an effective solution for this problem. The experimental results show that with four sensor observations (s1-s4), the proposed method has $33.38\%$ improvement on the accuracy of the simulation data around sensors s5-s8. If we consider the whole thermal map (including the region around sensors s1-s4), the improvement can be even larger.

We need to set three parameters for the proposed method: balance factor, magnitude variance, and distance variance.
The balance factor controls the magnitude of the adjustment for the simulated thermal map,
but the upper bound of this magnitude is determined by the sensing locations. The distance variance controls the scope of the regions on which an isolated sensor observation has significant influence. A small distance variance can lead to very good local calibration while a large one can provide global influence for an isolated sensor observation. The magnitude variance controls the influence of the CFD results and scales the influence of the distance variance. Reducing the magnitude variance can improve the local calibration but has no significant help for global calibration. If enough sensor observations are available we can set both small magnitude variance and small distance variance. If the number of available sensor observations is very limited, we should set relatively larger values for the two variances to achieve global performance.

In addition, the capacity and the performance of the proposed method is closely related to the number of sensor observations and the sensing locations. The issue of how many sensors are required and where to place them are very interesting topics for future research.

\section*{Acknowledgements}
This work was supported by Singapore's National Research Foundation under NRF-CRP8-2011-03 and partially supported by the Energy Research Institute at NTU (ERI@N).
The first author would like to thank Mr. Ke Ji, Mr. Xingyu Zhang, and Mr. Huynh Nam Khoa for their help in the experiments.
\bibliographystyle{IEEEtran}

\begin{thebibliography}{99}
\bibitem{zhai2006application}
Z.~Zhai, ``Application of computational fluid dynamics in building design:
  aspects and trends,'' \emph{Indoor and Built Environment}, vol.~15, no.~4,
  pp. 305--313, 2006.

\bibitem{chen2009ventilation}
Q.~Chen, ``Ventilation performance prediction for buildings: A method overview
  and recent applications,'' \emph{Building and Environment}, vol.~44, no.~4,
  pp. 848--858, 2009.

\bibitem{srebric2008cfd}
J.~Srebric, V.~Vukovic, G.~He, and X.~Yang, ``CFD boundary conditions for
  contaminant dispersion, heat transfer and airflow simulations around human
  occupants in indoor environments,'' \emph{Building and Environment}, vol.~43,
  no.~3, pp. 294--303, 2008.

\bibitem{hajdukiewicz2013formal}
M.~Hajdukiewicz, M.~Geron, and M.~M. Keane, ``Formal calibration methodology
  for CFD models of naturally ventilated indoor environments,'' \emph{Building
  and Environment}, vol.~59, pp. 290--302, 2013.

\bibitem{guillas2014bayesian}
S.~Guillas, N.~Glover, and L.~Malki-Epshtein, ``Bayesian calibration of the
  constants of the k--$\varepsilon$ turbulence model for a CFD model of street
  canyon flow,'' \emph{Computer Methods in Applied Mechanics and Engineering},
  vol. 279, pp. 536--553, 2014.

\bibitem{kajero2016kriging}
O.~T. Kajero, R.~B. Thorpe, T.~Chen, B.~Wang, and Y.~Yao, ``Kriging meta-model
  assisted calibration of computational fluid dynamics models,'' \emph{AIChE
  Journal}, 2016.

\bibitem{an2008appprop}
X.~An and F.~Pellacini, ``Appprop: all-pairs appearance-space edit
  propagation,'' in \emph{ACM Transactions on Graphics (TOG)}, vol.~27,
  no.~3.\hskip 1em plus 0.5em minus 0.4em\relax ACM, 2008, p.~40.

\bibitem{xu2013sparse}
L.~Xu, Q.~Yan, and J.~Jia, ``A sparse control model for image and video
  editing,'' \emph{ACM Transactions on Graphics (TOG)}, vol.~32, no.~6, p. 197,
  2013.

\bibitem{golub2012matrix}
G.~H. Golub and C.~F. Van~Loan, \emph{Matrix computations}.\hskip 1em plus
  0.5em minus 0.4em\relax JHU Press, 2013.

\end{thebibliography}

\end{document}